\documentclass[bibyear]{aa}
\usepackage{graphicx}
\usepackage{txfonts}

\begin{document}

\title{Magnetic fields and star formation in low-mass Magellanic-type and peculiar
galaxies\thanks{Based on observations with the 100-m telescope at Effelsberg
operated by the Max-Planck-Institut f\"ur Radioastronomie (MPIfR) on behalf
of the Max-Planck-Gesellschaft.
}
}

\author{W.~Jurusik\inst{\ref{inst1}\thanks{\email{jurusik@oa.uj.edu.pl}}}
\and R.~T.~Drzazga\inst{\ref{inst1}}
\and M.~Jableka\inst{\ref{inst1}}
\and K.~T.~Chy\.zy\inst{\ref{inst1}}
\and R.~Beck\inst{\ref{inst2}}
\and U.~Klein\inst{\ref{inst3}}
\and M.~We\.zgowiec\inst{\ref{inst4}}
}

\institute{Obserwatorium Astronomiczne Uniwersytetu
Jagiello\'nskiego, ul. Orla 171, 30-244 Krak\'ow, Poland\label{inst1}
\and Max-Planck-Institut f\"ur Radioastronomie, Auf dem H\"ugel 69, 53121 Bonn, Germany\label{inst2}
\and Argelander-Instit\"ut f\"ur Astronomie, Auf dem H\"ugel 71, 53121 Bonn, Germany\label{inst3}
\and Astronomisches Institut der Ruhr-Universit\"at Bochum, Universit\"atsstrasse 150, 44780 Bochum, Germany\label{inst4}
}

\date{Received 15 November 2013  / Accepted 6 June 2014}
\titlerunning{Magnetic fields and star formation in low-mass galaxies}
\authorrunning{W. Jurusik et al.}

\abstract
{}
{
We investigate how magnetic properties of Magellanic-type and perturbed objects
are related to star-forming activity, galactic type, and mass.
}
{
We present radio and magnetic properties of five Magellanic-type and two peculiar
low-mass galaxies observed at 4.85 and/or 8.35\,GHz with the Effelsberg 100-m telescope.
The sample is extended to 17 objects by including five
Magellanic-type galaxies and five dwarf ones.}
{
The distribution of the observed radio emission of low-mass galaxies at 4.85/8.35\,GHz is closely connected with the galactic optical discs, which are independent for unperturbed galaxies and those which show signs of tidal interactions. The strengths of total magnetic field are within 5-9\,$\mu$G, while the ordered fields reach 1-2\,$\mu$G, and both these values are larger than in typical dwarf galaxies and lower than in spirals. The magnetic field strengths in the extended sample of 17 low-mass galaxies are well correlated with the surface density of star formation rate (correlation coefficient of 0.87) and manifest a power-law relation with an exponent of $0.25\pm 0.02$ extending a similar relation found for dwarf galaxies. We claim that the production of magnetic energy per supernova event is very similar for all the various galaxies. It constitutes about 3\% ($10^{49}$\,erg) of the individual supernovae energy release. We show that the total magnetic field energy in galaxies is almost linearly related to the galactic gas mass, which indicates equipartition of the magnetic energy and the turbulent kinetic energy of the interstellar medium.  The Magellanic-type galaxies fit very well with the radio-infrared relation constructed for surface brightness of galaxies of various types, including bright spirals and interacting objects (with a slope of $0.96\pm0.03$ and correlation coefficient of 0.95). We found that the typical far-infrared relation based on luminosity of galaxies is tighter and steeper but more likely to inherit a partial correlation from a tendency that larger objects are also more luminous.
}
{The estimated values of thermal fractions, radio spectral indices, and magnetic field strengths
of the Magellanic-type galaxies are between the values determined for grand-design spirals and dwarf
galaxies. The confirmed magnetic field--star formation and radio--infrared relations
for low-mass galaxies point to similar physical processes that must be at work in all
galaxies. More massive, larger galaxies have usually stronger magnetic fields and
larger global star formation rates, but we show that their values of magnetic energy
release per supernova explosion are still similar to those of dwarf galaxies.
}

\keywords{Galaxies: general -- galaxies:  magnetic fields -- galaxies: interactions -- radio continuum: galaxies}

\maketitle

\section{Introduction}
\label{s:intro}

While galactic magnetism has been known for over 30 years and observed for various galaxy
types (e.g. Beck \& Wielebinski \cite{beck13}), it is still far from being fully understood.
Of particular importance are investigations concerning the generation and evolution
of magnetic fields in low-mass objects, in which spiral density waves might be too weak
to trigger star formation and influence magnetic fields.
In such galaxies, due to their lower gravitational potential, other processes, such as galactic
winds and stochastically propagating star formation, play a more important role.
For these objects, we do not know  the exact conditions required for the efficient
amplification of the large-scale magnetic fields and the relations of magnetic fields
to other phases of the interstellar medium (ISM) yet.

Radio observations and magnetic studies of low-mass galaxies are difficult
and thus rare to date.  Single-dish observations are preferable at centimetre wavelengths
as interferometric measurements of dwarfs may miss their low-level extended emission due to the missing
spacings (e.g. Heesen et al. \cite{heesen11}). Even with the largest single radio dishes
observations of low-mass galaxies are challenging.  For example, Chy\.zy et al.~(\cite{chyzy11})
performed a systematic survey of dwarf galaxies in the Local Group. However, only 3 out of 12 dwarfs
were radio-detected in sensitive observations with the Effelsberg 100-m telescope at 2.64\,GHz.
The results obtained indicated that the magnetic fields within the Local Group
dwarfs are rather weak with a mean value of the total field strength of only $4.2\,\mu$G.
A stronger magnetic field was only observed in the starbursting dwarf IC\,10.
The radio undetected dwarfs were of the lowest total mass and global star formation rate (SFR).
Based on seven radio-detected low-mass dwarfs (from the Local Group and outside it),
a power-law relation of the magnetic field strength and the surface density of star formation
rate ($\Sigma$\,SFR) with an index of $0.30\pm 0.04$ was proposed.

To extend that study and investigate the relations observed for dwarf galaxies,
 we report results from new radio polarimetric observations of seven low-mass galaxies
of Magellanic-type or peculiar morphology in this paper. The sizes and masses of these objects
are between dwarf and typical spiral galaxies. Consequently, such galaxies have a better
chance to be radio-detected than dwarfs of the Local Group and hence are suitable for
further investigation of the relations found in dwarf objects (Chy\.zy et al.~\cite{chyzy11}).
In the extended sample of low-mass galaxies based on our new observations and on the
published ones, we analyse the properties of total radio and polarised emissions, construct
a radio--infrared correlation diagram, and investigate a generation of magnetic fields
in response to the observed star formation activity. We investigate the net production
of magnetic fields per single supernova explosion and relation
of magnetic energy to global properties of galaxies.

\begin{table*}[t]
\caption{Basic properties of the observed galaxies.}
\centering
\begin{tabular}{lccccccr}
\hline\hline
Name	&	Type\tablefootmark{a}&Inclination\tablefootmark{b}&Size\tablefootmark{a}     & Distance\tablefootmark{c}&	${\mathrm{v_{rot}}}$\tablefootmark{b}	&	$M_{\mathrm{HI}}$\tablefootmark{d} &	$M_{\mathrm{tot}}$\\
        &		&$[^\circ]$	&[']	                &	Mpc	& km\,s$^{-1}$	        &	$10^8\,\mathrm{M}_{\sun}$ &	$10^9\,\mathrm{M}_{\sun}$\\
\hline
NGC2976	&	SAc pec	&	60.5	&	$5.9\times2.7$	&	3.6	&	$58.5\pm2.4$	&	1.5	\tablefootmark{g}  & 2.4\\
NGC3239	&	IB(s)m pec&	46.8	&	$5.0\times3.3$	&	8.3	&	$95.0\pm2.5$	&	13.0	                   & 12.4\\
NGC4027 &	SB(s)dm	&	42.3	&	$3.2\times2.4$	&	15.2\tablefootmark{e}&	$97.8\pm4.8$	&	40.0\tablefootmark{h} & 15.5\\
NGC4605	&	SB(s)c pec&	70.0	&	$5.8\times2.2$	&	5.5	&	$60.9\pm2.0$	&	2.0                        & 3.9\\
NGC4618	&	SB(rs)m	&	57.6	&	$4.2\times3.4$	&	7.8	&	$65.7\pm4.1$	&	11.0	                   & 4.7\\
NGC5204	&	SA(s)m	&	58.8	&	$5.0\times3.0$	&	4.7	&	$55.9\pm1.1$	&	6.3	                   & 2.4\\
UGC11861&	SAB(s)dm&	75.0	&	$3.5\times2.6$	&	20.9\tablefootmark{f}&	$114.6\pm3.1$	&	87.0               & 31.8\\
\hline
\end{tabular}
\tablefoot{
Data taken from
\tablefoottext{a}{NED}
\tablefoottext{b}{LEDA}
\tablefoottext{c}{Kennicutt et al. (\cite{kennicutt08})}
\tablefoottext{d}{Tully et al. (\cite{tully88})}
\tablefoottext{e}{Bell et al. (\cite{bell03})}
\tablefoottext{f}{James et al. (\cite{james04})}
\tablefoottext{g}{Stil \& Israel (\cite{stil02})}
\tablefoottext{h}{Bottinelli et al. (\cite{bottinelli82})}
}
\label{t:sample}
\end{table*}

\section{Sample selection and data reduction}
\label{s:sample}

For the radio polarimetric observations, we selected five Magellanic-type 
galaxies showing large apparent optical discs ($>3^\prime$) of various morphology.
All of them have \ion{H}{i} masses larger than the LMC. To provide a better
coverage of galactic masses as well as a closer correspondence with the dwarf galaxies, we supplemented
the sample with two peculiar galaxies (NGC\,2976 and NGC\,4605) with \ion{H}{i} masses that are
smaller than the LMC. The main properties of all the seven objects are summarised in Table \ref{t:sample}.
For the sake of comparison, we present not only the \ion{H}{i} mass of galaxies but also their ``indicative'' total masses 
estimated according to the formulae: $M_{\mathrm{tot}}=0.5\,G^{-1}\,A\,{\mathrm v_{rot}}^2$,
where $A$ is the linear extent of the galactic major axis based on optical observations,
${\mathrm v_{rot}}$ is the maximum velocity rotation corrected for inclination, and $G$ is the
gravitational constant (cf. Karachentsev et al. \cite{karachentsev04}).
Values of $A$ (in arcmin) and ${\mathrm v_{rot}}$ are also given in Table \ref{t:sample}.
We note that the 3D mass distribution, including the dark matter is not known for these objects and
the ``indicative'' mass is the first order approximation of the total dynamical mass of these
galaxies by assuming a spherical mass distribution.

Some galaxies in the sample have properties similar to those of spirals, for example, having an \ion{H}{i} mass
above $10^9\,\mathrm{M}_{\sun}$ (NGC\,4027) or rotational velocity larger than
$100$\,km\,s$^{-1}$ (e.g. UGC\,11861), while others have rotational
velocities and sizes similar to those of dwarf irregulars (as e.g. NGC\,5404 and NGC\,2976).
The total  ``indicative'' mass is largest for UGC\,11861 ($3.18\times 10^{10}\,\mathrm{M}_{\sun}$), which is
still one order of magnitude smaller than the mass of the Andromeda galaxy
($3.20\times 10^{11}\,\mathrm{M}_{\sun}$), when estimated in the same way.
Our objects are at different stages of gravitational interaction: from
galaxies that do not have distinct morphological distortions, like UGC\,11861, and weakly
interacting (NGC\,4618) to disturbed ones (NGC\,3239, see Sect. \ref{s:radio} for details).

\begin{table*}[t]
\caption{Parameters of observations, integrated radio data, and thermal fraction of studied galaxies
at 4.85\,GHz.
}
\centering
\begin{tabular}{lccccccc}
\hline\hline
Galaxy & Map No. & $\sigma^{TP}_{4.85}$ & $\sigma^{PI}_{4.85}$ & $TP_{4.85}$ & $PI_{4.85}$  & $f_{\rm th, 4.85}$ \\ 
       & 4.85 GHz& mJy/b.a.           & mJy/b.a.           & mJy         & mJy              &             \\
\hline
NGC2976 & 14     & 0.6                & 0.1                & $32.0\pm2.5$& $1.39\pm0.30$    & $0.31\pm0.07$\\
NGC3239 & 12     & 0.8                & 0.1               & $16.5\pm0.9$& $0.22\pm0.14$     & $0.34\pm0.07$\\
NGC4027 & 11     & 1.5                & 0.2                & $40.7\pm3.2$& $0.70\pm0.39$    & $0.29\pm0.06$\\
NGC4605 & 6      & 0.5                & 0.7                & $37.6\pm2.5$& $<1.40$          & $0.23\pm0.05$\\
NGC4618 & 17     & 0.5                & 0.1                & $15.0\pm1.2$& $0.49\pm0.14$    & $0.29\pm0.06$\\
NGC5204 & $-$    & $-$                & $-$                & $-$         & $-$              & $-$\\
UGC11861& 8      & 1.5                & 0.1                & $15.1\pm1.5$& $0.53\pm0.17$    & $0.11\pm0.02$\\
\hline
\end{tabular}
\label{t:radio4}
\end{table*}

\begin{table*}[t]
\caption{Parameters of observations, integrated radio data, and thermal fraction of studied galaxies
at 8.35\,GHz.
}
\centering
\begin{tabular}{lcccccccccc}
\hline\hline
Galaxy &  Map No. & $\sigma^{TP}_{8.35}$ & $\sigma^{PI}_{8.35}$ & $TP_{8.35}$ & $PI_{8.35}$  & $f_{\rm th,8.35}$ \\
       &  8.35 GHz& mJy/b.a.           & mJy/b.a.           & mJy         & mJy              &             \\
\hline
NGC2976 & $-$     & $-$                & $-$                & $-$         & $-$              & $-$\\
NGC3239 &  17      & 0.4                & 0.08                & $10.4\pm0.8$& $<0.2$ & $0.50\pm0.11$\\
NGC4027 & 6       & 0.7                & 0.1                & $23.7\pm1.1$& $<0.2$ & $0.47\pm0.10$\\
NGC4605 & $-$     & $-$                & $-$                & $-$         & $-$              & $-$\\
NGC4618 & 16      & 0.3                & 0.08                & $13.6\pm0.9$& $<0.3$ & $0.30\pm0.06$\\
NGC5204 & 4       & 0.5                & 0.1                & $8.2\pm0.8$ & $<0.3$ & $0.41\pm0.09$\\
UGC11861& $-$                & $-$                & $-$         & $-$      & $-$\\
\hline
\end{tabular}
\label{t:radio8}
\end{table*}

Our radio polarimetric observations were performed with the Effelsberg 100-m telescope.
All seven galaxies were observed at 4.85\,GHz with the dual-horn system (with the horn separation of 8\arcmin),
and a correlation polarimeter in the backend by offering an output of Stokes parameters I, Q, and U.
The individual coverages were obtained in both azimuth and elevation directions.
To establish a flux density scale, we observed the calibration source 3C\,286 and
assumed its total flux as 7.45\,Jy, according to the flux scale of Baars et al. (\cite{baars77}).

The data were reduced in NOD2 package, which  combined the coverages using software
beam-switching (Emerson et al. \cite{emerson79}) and spatial-frequency weighting methods (Emerson \&
Gr\"ave~\cite{emerson88}). The maps were then digitally filtered to remove the
spatial frequencies that correspond to noisy structures that are smaller than the telescope beam of $2.5\arcmin$.
The details of the observations are given in Tables \ref{t:radio4} and \ref{t:radio8}.
In the case of NGC\,5204, the processing of the data did not bring satisfactory results due to
strong artefacts of weather origin and a number of background sources
that made it impossible to properly fit baselines in individual coverages. Therefore, the maps of
NGC\,5204 at 4.85\,GHz were discarded in the analysis.

A subsample of four galaxies (NGC\,3239, NGC\,4027, NGC\,4618, and NGC\,4204) was
observed at 8.35\,GHz as well. At this frequency, the single-horn receiver in the secondary focus was
used. High sensitivity of the  observations was guaranteed by using a frequency bandwidth of 1\,GHz.
The objects were scanned along the RA and Dec directions. The calibrator 3C\,286 was
once again applied for the flux scale calibration, assuming 5.22\,Jy total flux (Baars et al. \cite{baars77}).
We then combined individual coverages through a procedure similar to 4.85\,GHz by obtaining the
final maps in Stokes I, Q, and U with a resolution of $1.4\arcmin$.

The final maps from both frequencies were converted to FITS format and further analysed in the AIPS
program. The maps of polarised intensity, polarisation degree and polarisation position
angles were thus obtained, and the r.m.s (root mean squared) noise levels were determined  (Tables \ref{t:radio4}, \ref{t:radio8}).
The polarised intensity was properly corrected for the noise distribution of Q and U signal.

\section{Results}
\label{s:results}

\begin{figure*}
\begin{minipage}[b]{1\linewidth}
\centering
\includegraphics[clip,angle=0,width=7cm]{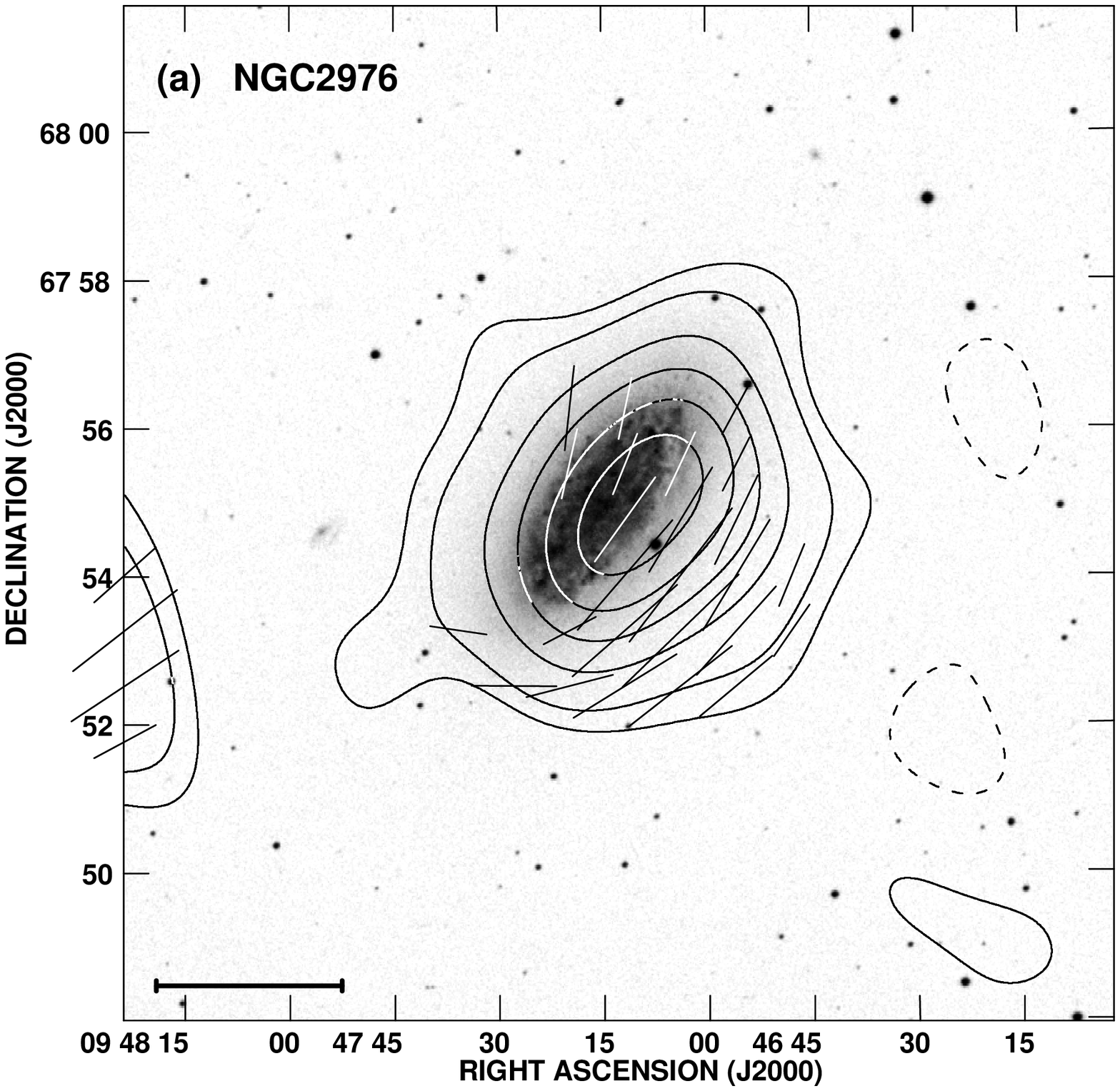}
\includegraphics[clip,angle=0,width=7cm]{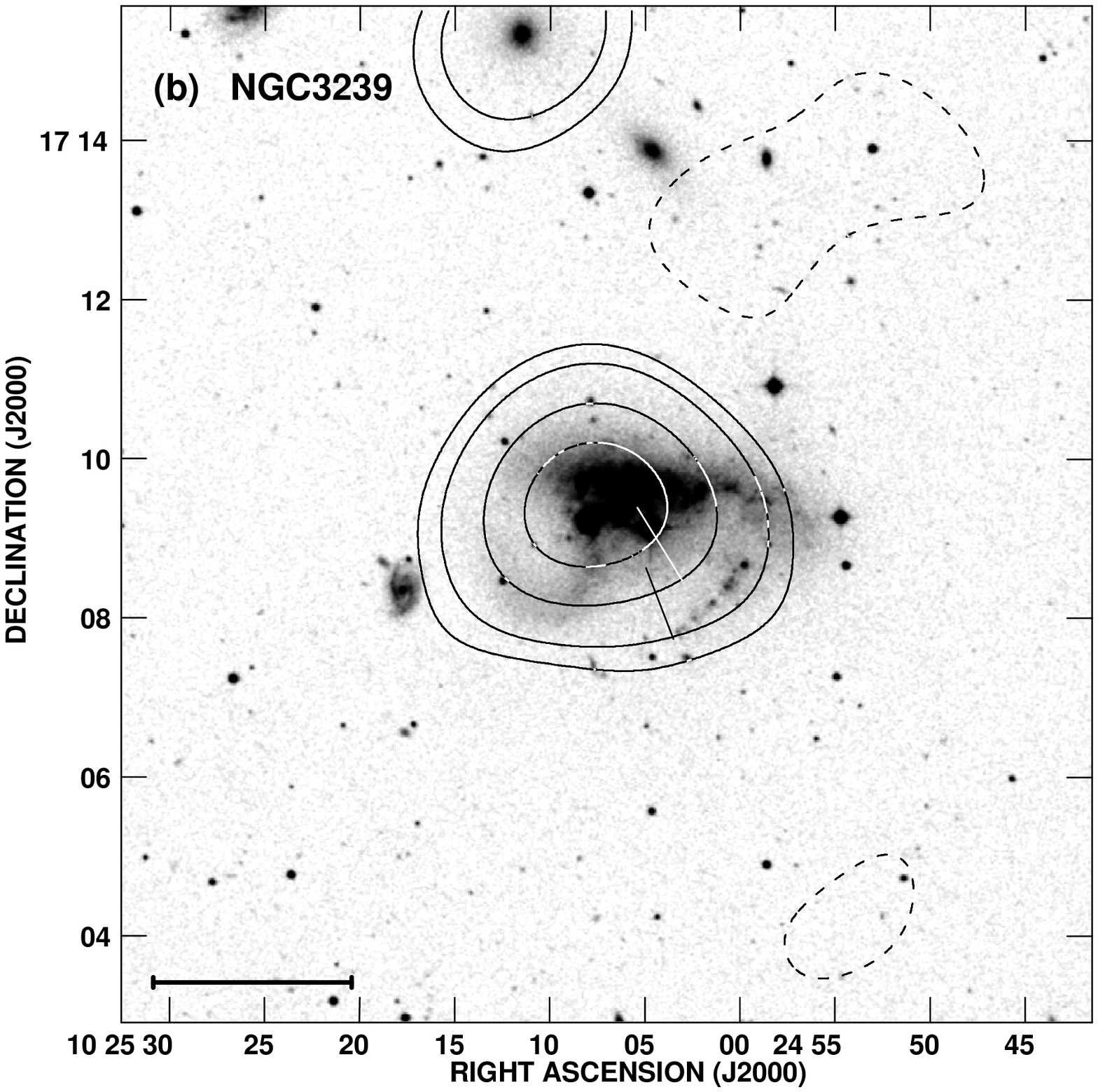}
\end{minipage}\\
\begin{minipage}[b]{1\textwidth}
\centering
\includegraphics[clip,angle=0,width=7cm]{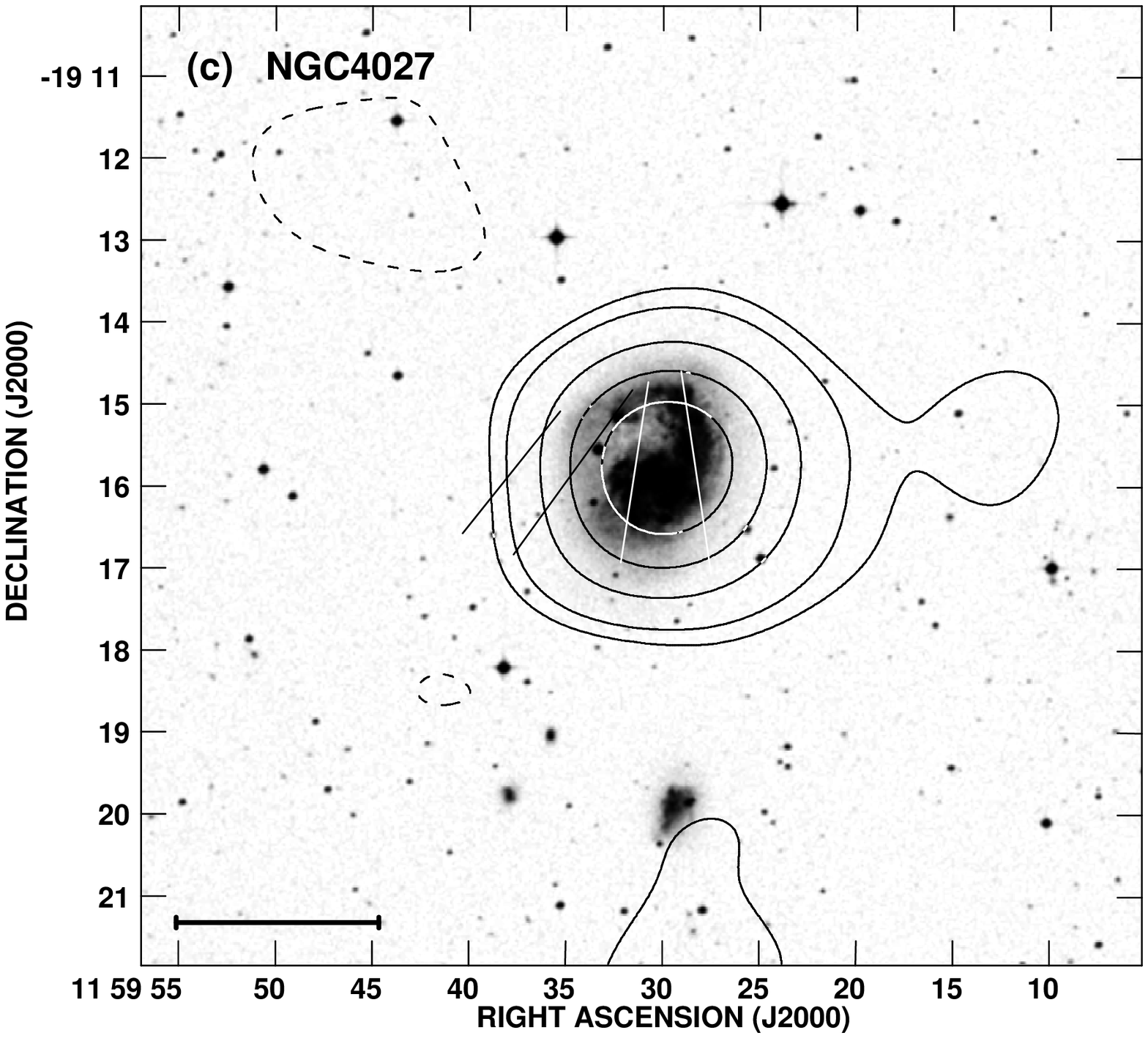}
\includegraphics[clip,angle=0,width=6.8cm]{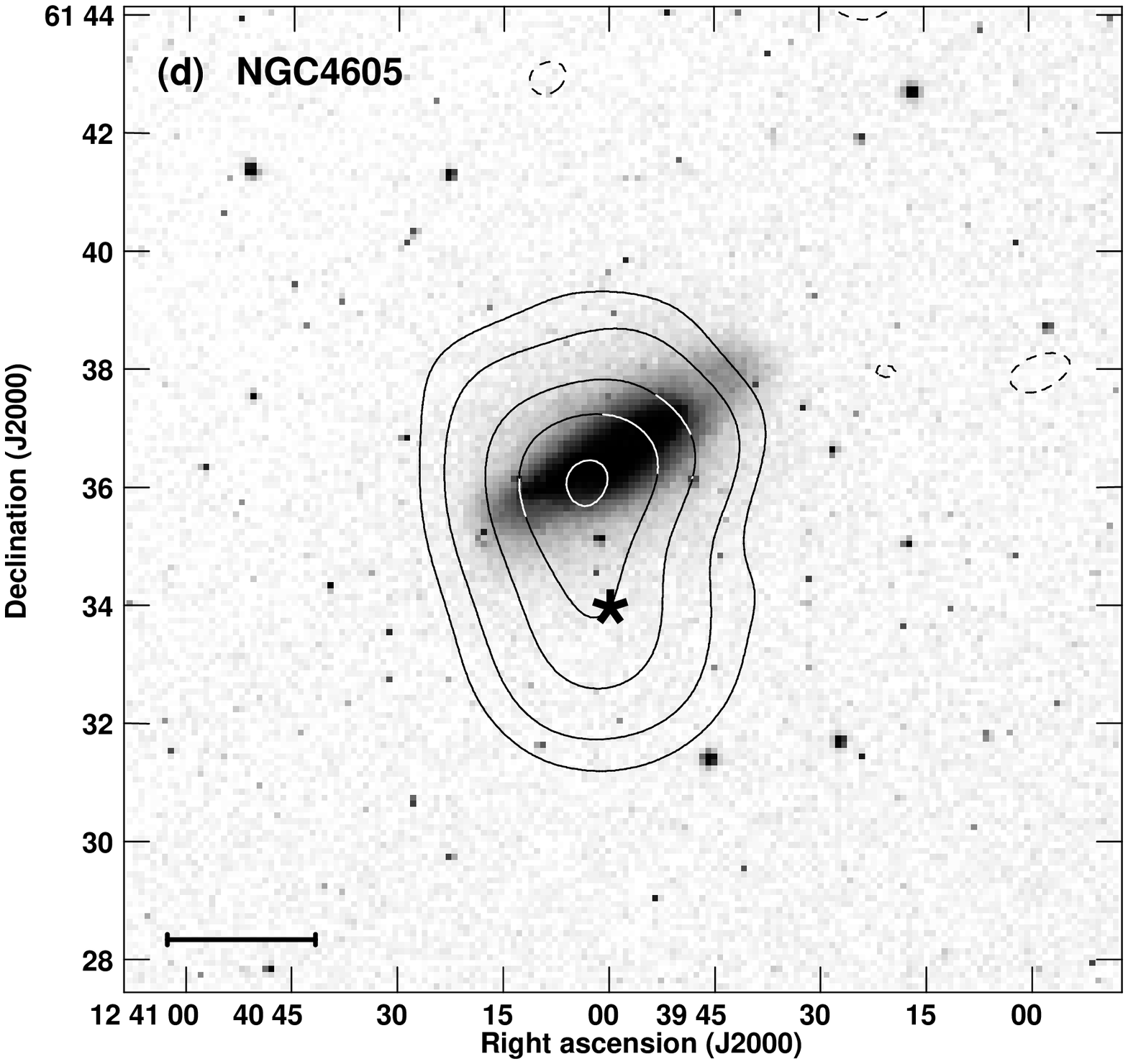}
\end{minipage}\\
\begin{minipage}[b]{1\textwidth}
\centering
\includegraphics[clip,angle=0,width=7cm]{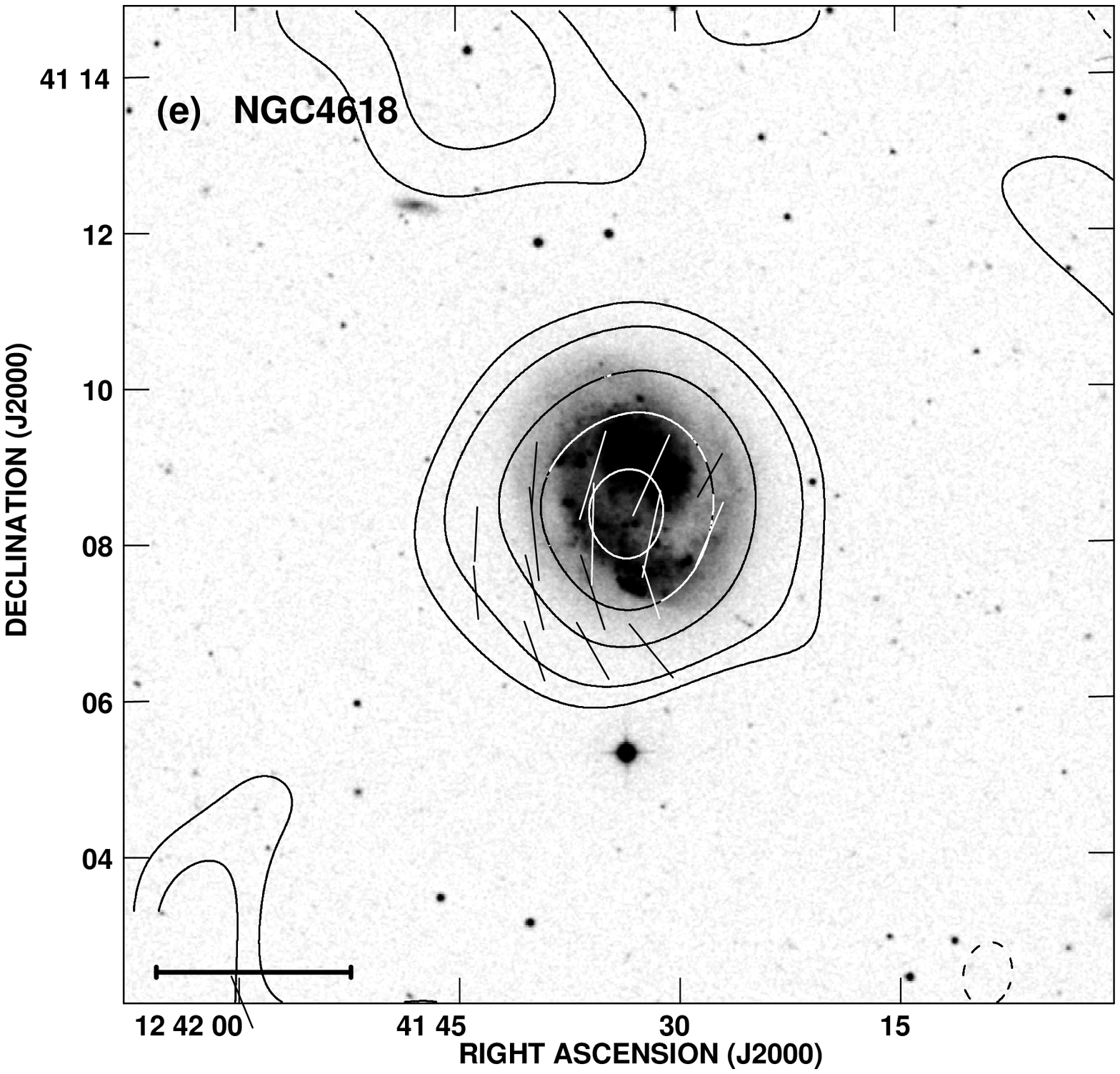}
\includegraphics[clip,angle=0,width=7cm]{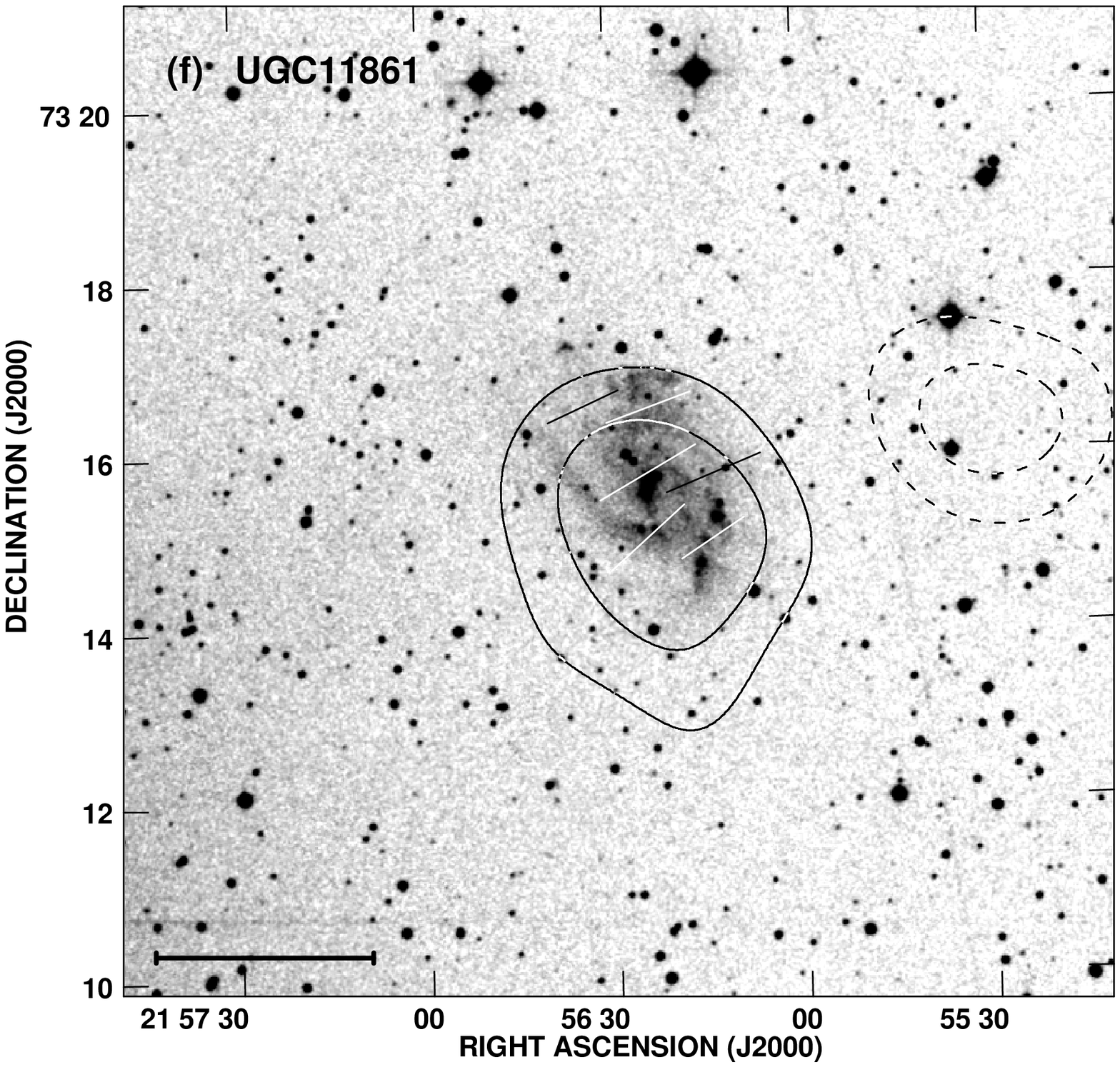}
\caption{Total-power contours at 4.85\,GHz and B-vectors of polarised intensity of
NGC\,2976\,(a), NGC\,3239\,(b), NGC\,4027\,(c), NGC\,4605\,(d), NGC\,4618\,(e),
UGC\,11861\,(f) superimposed on the DSS blue images. The contour levels are as
follows: 0.6 (a), 0.8 (b), 1.5 (c), 0.5 (e), 1.5 (f) $\times$ (-5, -3,
3, 5, 10, 15, 20, 25)\,mJy/beam and 0.45 $\times$ (-3, 3, 10, 30, 50, 70)\,mJy/beam (d).
The map resolution is 2.5$'$\,HPBW, as shown as a horizontal bar at bottom left. For all
the images, a vector of 1$'$ length corresponds to a polarised intensity of
0.3\,mJy/beam. The position of a background source subtracted from emission of
NGC\,4605 (d) is marked by an asterisk (Sect. \ref{s:results}).}
\label{f:6cm}
\end{minipage}
\end{figure*}

\begin{figure*} [t]
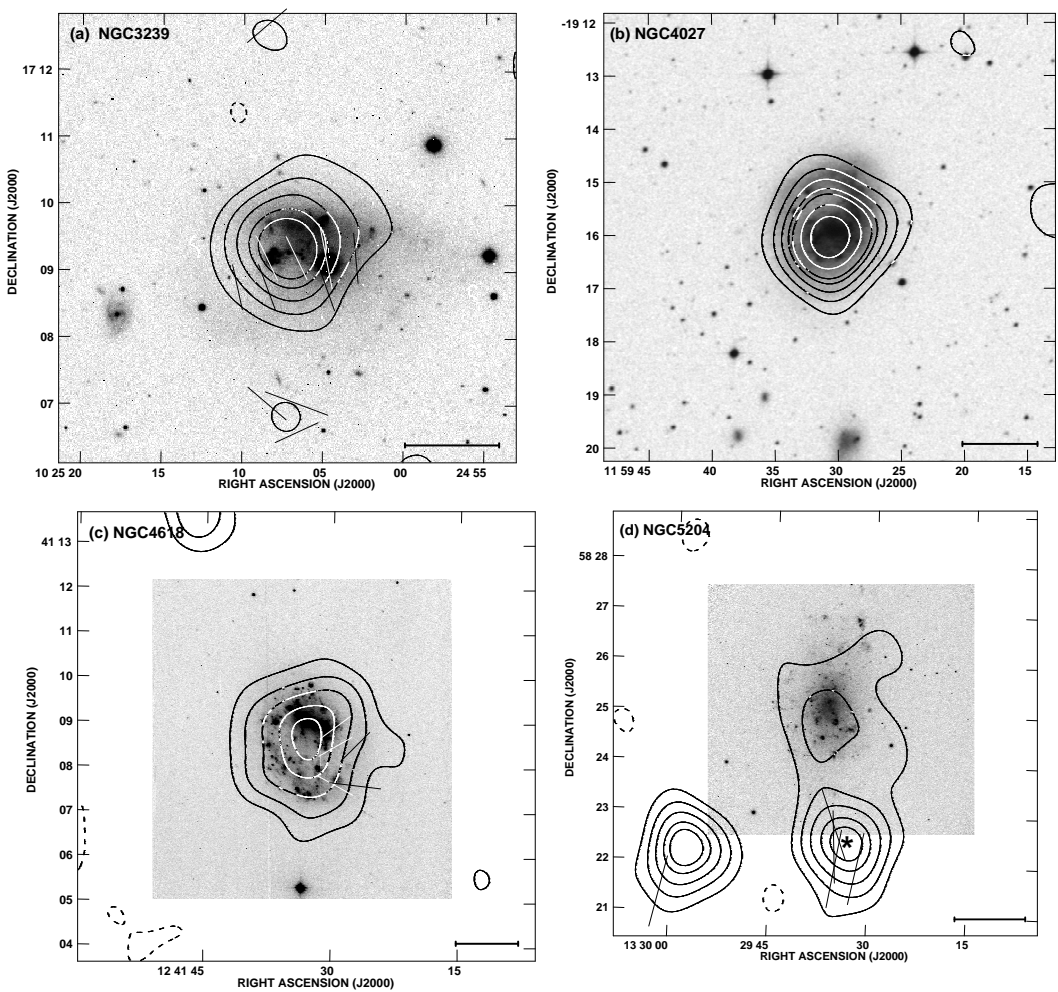

\begin{minipage}[b]{1\linewidth}
\centering
\includegraphics[clip,angle=-90,width=7cm]{3239_3ha.eps}
\includegraphics[clip,angle=-90,width=7cm]{4027_3DSS.eps}
\end{minipage}\\
\begin{minipage}[b]{1\textwidth}
\centering
\includegraphics[clip,angle=-90,width=7cm]{4618_3ha.eps}
\includegraphics[clip,angle=-90,width=6.5cm]{5204_3ha.eps}
\caption{Total-power contours at 8.35\,GHz and B-vectors of polarised
intensity of NGC\,3239 (a), NGC\,4618 (c), NGC\,5204 (d),
superimposed on the H$\alpha$ images and  NGC\,4027 (b) superimposed
on the DSS blue image. The contour levels are as follows: 0.5 (a),
0.7 (b), 0.3 (c), 0.45 (d), $\times$ (-5, -3, 3, 6, 9, 12, 14, 18,
22)\,mJy/beam. The map resolution is 1.4$'$\,HPBW, as shown as a
horizontal bar at bottom right.
For all the images, a vector of 1$'$ length corresponds to the polarised
intensity of 0.3\,mJy/beam. The H$\alpha$ images of NGC\,3239 and
NGC\,5204 are from NED and those for NGC\,4618 from Knapen et al. (\cite{knapen04}).
}
\label{f:3cm}
\end{minipage}
\end{figure*}

\begin{figure*} [t]
\begin{minipage}[b]{1\linewidth}
\centering
\includegraphics[clip,angle=0,width=6.8cm]{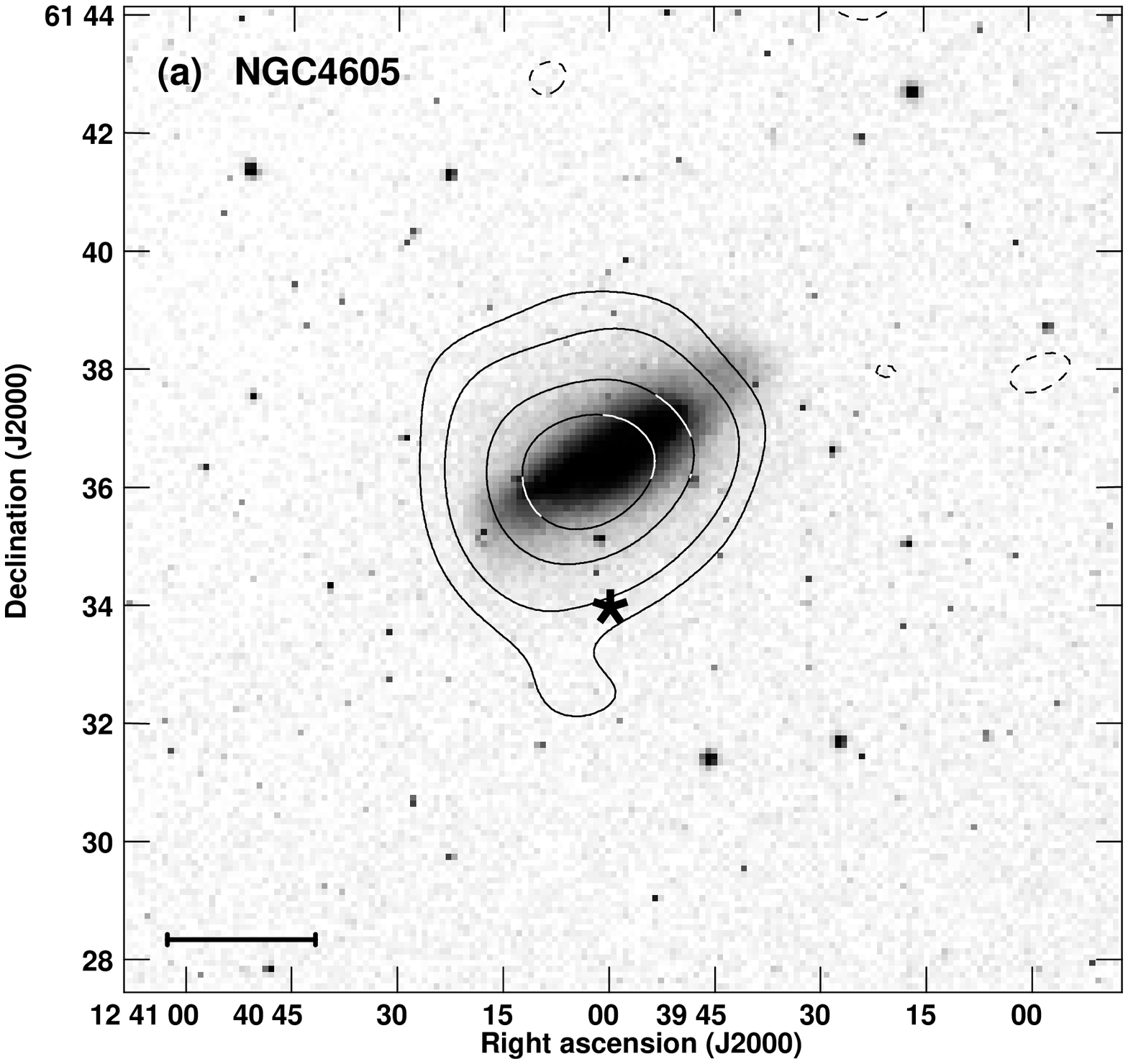}
\includegraphics[clip,angle=0,width=6.8cm]{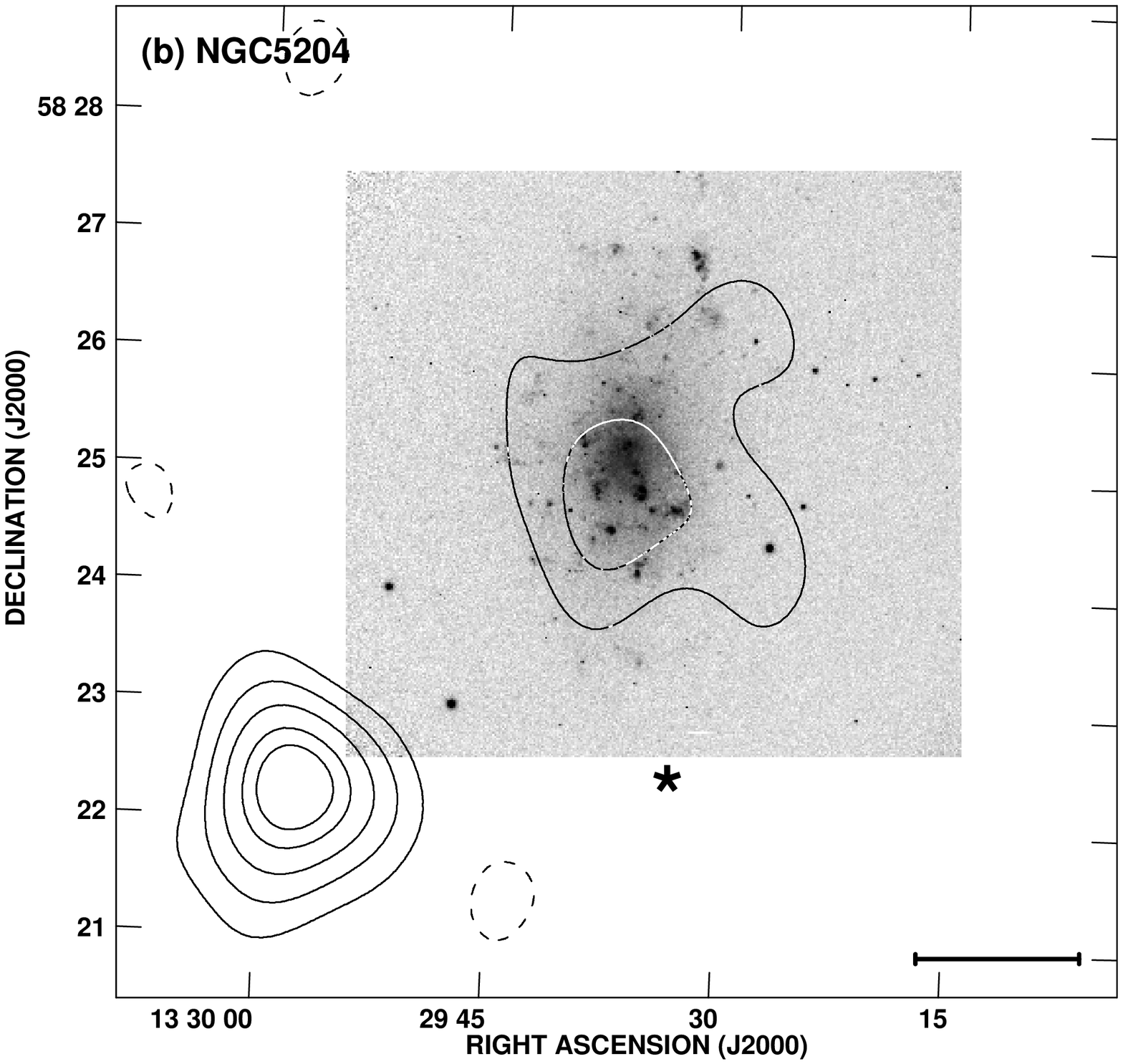}
\caption{
Total-power contours of radio emission without nearby background sources
for NGC\,4605 at 4.85\,GHz (a) and NGC\,5204 at 8.35\,GHz\,(b).
The contour levels and other map parameters are the same as in Figs. \ref{f:6cm} and
\ref{f:3cm}. The positions of the subtracted background sources are marked by
asterisks.}
\label{f:sub}
\end{minipage}
\end{figure*}

\subsection{Properties of radio emission}
\label{s:radio}

The distribution of radio emission at 4.85\,GHz and 8.35\,GHz and the structure
of magnetic fields for all galaxies are shown in Figures \ref{f:6cm} and
\ref{f:3cm}, respectively. The lengths of the presented B-vectors
are proportional to the polarised emission, and their orientation results from
the observed E-vectors rotated by $90\degr$.
We do not apply any corrections for Faraday effects: although a typical rotation
measure in spiral galaxies is up to $\pm 100\,{\rm rad\,m}^{-2}$ (e.g. We\.zgowiec
et al. \cite{wezgowiec07}), the corresponding rotation of the polarisation plane
is up to $23\degr$ at 4.85\,GHz and $7\degr$ at 8.35\,GHz. In case of our sample of
galaxies with a regular magnetic field strength less than $2\,\mu$G (Sect. \ref{s:magnetic}),
a typical density of thermal electrons in the galactic disc of
$n_e=0.03\,{\rm cm}^{-3}$, and a pathlength of 500\,pc, the resulting rotation measure
will be even smaller (about $24\,{\rm rad\, m}^{-2}$), and hence, the
E-vectors are expected to rotate up to about $6\degr$ at 4.85\,GHz.

The general radio properties of our galaxies and their relations to morphologies as
seen in other ISM components are as follows:

{\em NGC\,2976.} --- This peculiar late-type galaxy that belongs to
the M\,81 group of galaxies is the smallest object in our sample
(of a diameter of 6.2\,kpc). In optical images, there is just a ``pure disc'' morphology
with a sharp edge and without any bulge or spiral structure.
Our 4.85\,GHz data show asymmetric radio emission on  both sides of the optical
disc with an extension to the south-west (Fig. \ref{f:6cm}a). The polarised
emission is even more asymmetric with the peak shifted off the
optical disc to the south. The magnetic field B-vectors
can be seen outside the galactic disc as parallel to its southern edge. Since
the outer parts of NGC\,2976 were probably undisturbed for a long time
(Bronkalla et al.  \cite{bronkalla92}), this configuration of the ordered
field can indicate that the magnetic field was stretched or compressed
by external forces acting long ago. We will analyse higher
resolution VLA data for this galaxy in a future paper
(Drzazga et al. in preparation).

{\em NGC\,3239.} --- This galaxy is gravitationally bound with six
other galaxies, forming a group (Huchra \& Geller \cite{huchra82}).
The H$\alpha$ distribution is patchy (Fig. \ref{f:3cm}a), showing a
possible remnant of a disc structure and (probably tidal) tails.
The present \ion{H}{II} luminosity function is very similar to
the LMC, including a supergiant \ion{H}{II} region with the
luminosity of 76\% of 30\,Dorarus (Krienke \& Hodge \cite{krienke91}).
The radio emission at 4.85\,GHz corresponds well to the optical disc (Fig.
\ref{f:6cm}b). Some very weak polarised emission at $3\sigma$ level
($0.24\mu$\,Jy/b.a.) is also detected between the centre and the southern
optical tail. At 8.35\,GHz, the strongest radio emission is associated with the brightest
\ion{H}{II} regions (Fig.~\ref{f:3cm}a). Only the upper limit of
the emission could be estimated in polarisation (Table  \ref{t:radio8}).

{\em NGC\,4027.} --- This is a peculiar, asymmetric late-type spiral
with a small, separated companion (at RA$_{2000}$=$\rm
11^h59^m29.4^s$, Dec$_{2000}$=$-19\degr19\arcmin 55\arcsec$).
The galaxy has no bulge, and the nucleus is embedded in a bar (Eskridge et al.
\cite{eskridge02}). Its peculiar morphology and optical asymmetry is due to
one of its spiral arms being bright and massive (Figs.~\ref{f:6cm}c).
The NGC\,4027 resides in a small group of galaxies however, there is no
unequivocal evidence of any encounter or merger effects.
With the low resolution of our maps at 4.85 and 8.35\,GHz, the radio emission covers
the whole galactic disc (Figs.~\ref{f:6cm}c and ~\ref{f:3cm}b, respectively). 
A patch of radio emission to the west, visible at both frequencies, has its counterpart
in the NVSS map at 1.4\,GHz. However, its physical relation with the
galaxy is not clear. Polarised emission at $3.5\sigma$ level is detected
at 4.85\,GHz (Table \ref{t:radio4}) but only an upper limit of polarisation
is obtained at 8.35\,GHz (Table \ref{t:radio8}).

{\em NGC\,4605.} --- The galaxy is located at the edge of M81/M82 group, and
its rigid-body rotation creates problems in the CDM modelling of this
object (Sofue \cite{sofue98}). Its peculiar morphology is similar
to NGC\,2976, which does not feature either a central component or a distinct spiral pattern.
The radio emission from the galactic disc is quite noticeable at 4.85\,GHz
(Fig. \ref{f:6cm}d),   especially in the map with the $20\,\mu$Jy background
source to the south subtracted (Fig. \ref{f:sub}a). However, the resolution of our observations did not
allow us to make a more detailed comparison with the optical data.
No polarised emission has been detected here.

{\em NGC\,4618.} --- This galaxy manifests evidence of
tidal interactions: there is one prominent spiral arm to the south and
a bar extending further to the southwest (Fig. \ref{f:6cm}e). The peak of total
radio emission at both 4.85 and 8.35\,GHz (Fig. \ref{f:3cm}c) is set off from
the brightest central region in the H$\alpha$ map, towards the southern spiral arm.
The \ion{H}{i} distribution (Kaczmarek \& Wilcots \cite{kaczmarek12}) is therefore
compatible with radio emission extending from the bar to the south,
which further joins a bright \ion{H}{i} ring encircling the entire
galaxy. The polarised radio emission at 4.85\,GHz is detected in the SE part,
where the B-vectors are oriented roughly along the southern optical spiral
arm and resemble other tidally perturbed galaxies (Drzazga et al. \cite{drzazga11}).
The degree of polarisation of about 3\% at 4.85\,GHz proves the presence
of ordered magnetic fields in the outer part of the galaxy.
At 8.35\,GHz, a spot of polarised signal at about 3\, r.m.s. noise level is
visible in the southwest part of the disc but not in the southeast part (Fig. \ref{f:3cm}c).
It is possible that the noise distribution at 8.35\,GHz is not uniform in this field:
this frequency is more sensitive to weather conditions and scanning effects than at
4.85\,GHz. Thus, we do not consider this spot as a proper detection of polarisation
at 8.35\,GHz.

{\em NGC\,5204.} --- The galaxy is located on the periphery of M\,101
group but has no close companion. The optical emission is symmetric,
and the radio emission at 8.35\,GHz corresponds to it (Figs. \ref{f:3cm}d, \ref{f:sub}b).
The UV emission from the galaxy is dominated by very massive stars
($>30\,\rm{M}_{\sun}$), which seems to be reflected in our data by
a high thermal fraction at 8.35\,GHz $f_{\rm th,8.35}=0.41$ (see Sect. \ref{s:magnetic} and
Table \ref{t:radio8}). A remarkable flattening of the radio spectrum
due to significant thermal component at frequencies above 10\,GHz was claimed by
Fabbiano \& Panagia (\cite{fabbiano82}).

{\em  UGC\,11861.} --- This isolated galaxy is the largest (with a diameter
of 18.2\,kpc) and the most massive ($M_{\mathrm{HI}}\approx 9\times 10^9\,\rm{M}_{\sun}$)
object in our sample.  It shows a weak bar and a few dispersed spiral arms
in optical images (Fig. \ref{f:6cm}f).
The radio emission does not manifest
any special features when observed with the resolution of $2.5\arcmin$
and corresponds roughly to the optical extent of the object.

In summary, our observations indicate a substantial correspondence
between radio emission and optical discs for all galaxies,
regardless if the galaxy is isolated or has signs of gravitational interactions
(NGC\,3239, NGC\,4027). Some visible shifts between maxima in radio and
optical distributions (e.g. NGC\,4618, Fig.~\ref{f:3cm}c) may arise from
more extended radio emission towards star-forming regions out of the main body
of galaxy discs and from lower resolution of the radio maps.
Polarised emission was detected for five out of six galaxies  observed
at 4.85\,GHz. At 8.35\,GHz, only upper ($2\sigma$) limits for the
observed 4 galaxies were possible to be estimated (Table \ref{t:radio8}).
The percentage of polarisation is low between 1\% and 4\% at 4.85\,GHz.
This is quite understandable when we keep in mind possible beam depolarisation
due to low resolution of our single-dish observations.
The low degree of polarisation may partly be intrinsic to the galaxies.

The complete determination of intrinsic configuration of magnetic fields in
low-mass galaxies, as well as testing for the presence of regular (unidirectional)
magnetic fields of a large-scale dynamo origin, which requires high resolution
information on Faraday rotation. There are no such data
for our galaxies to date. They can hopefully be provided by synthesis
observations (like with the JVLA) at centimetre wavelengths.

\subsection{Total fluxes and spectral indices}
\label{s:fluxes}

\begin{figure}
\includegraphics[angle=0,width=\columnwidth]{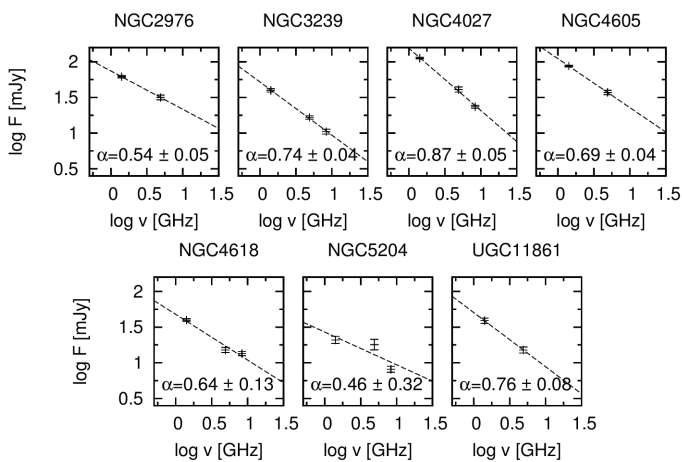}
\caption{Spectral indices for Magellanic-type and peculiar galaxies.
}
\label{f:spix}
\end{figure}

Our observations at 4.85\,GHz or 8.35\,GHz enabled detection of all seven galaxies and
measurements of their integrated fluxes in total emission and for a part of them in
the polarised emission as well (Tables \ref{t:radio4}, \ref{t:radio8}). Up to now, only
1.4\,GHz observations of all these objects were available from the VLA whole sky survey
(NVSS, Condon et al. \cite{condon98}). We applied a special procedure to our
radio maps to remove any possible influence of background sources located close
to our targets on their flux measurements. Accordingly, the galaxies' neighbourhood was
searched for background sources in higher resolution maps from the NVSS and FIRST
surveys. In two cases, (NGC\,4605 and NGC\,5204) such sources were actually found.
In their positions, point sources were subtracted from the maps, while their
fluxes were adjusted to obtain a smooth background distribution after the
subtraction (cf. Chy\.zy et al. \cite{chyzy03}). In the case of
NGC\,4605, a background source of 20\,mJy at 4.85\,GHz was
subtracted at the position of RA=$\rm 12^h39^m59.4^s$, Dec=$61\degr 33\arcmin 43\arcsec$.
From the map of NGC\,5204 at 8.35\,GHz a source of 8.0\,mJy was
subtracted at RA=$\rm 13^h29^m31.8^s$, Dec=$58\degr 22\arcmin 24\arcsec$.
The final estimated total fluxes ($TP$) of all galaxies are given in
Tables \ref{t:radio4} and \ref{t:radio8}. There was no need to remove background sources
from the polarisation ($PI$) maps.

Since all our galaxies can also be  found in the NVSS survey, we were able to
construct simple global spectra using our data and the measured NVSS
integrated fluxes. The two- or three-point spectra obtained in this way
are shown in Figure \ref{f:spix}. The estimated spectral index $\alpha$
($S_\nu \propto \nu^{-\alpha}$) is in the range from 0.46 (NGC\,5204)
to 0.87 (NGC\,4027) with an average of $0.67\pm0.14$.
These values are close to the typical spirals for which Gioia et al.
(\cite{gioia82}) derived a mean value of $0.74\pm0.12$. Low-mass dwarf galaxies
have typically flatter spectra and a global spectral index of around $0.38\pm0.20$
(Klein \& Gr\"ave \cite{klein86}). Therefore, our Magellanic-type and
perturbed galaxies resemble rather massive spirals than the low-mass dwarfs
and in the radio emission the synchrotron component apparently dominates over
the thermal one.

\subsection{Magnetic field strengths}
\label{s:magnetic}

\begin{table*}[t]
\caption{Magnetic field strength in $\mu$G for low-mass galaxies.
}
\centering
\begin{tabular}{lcccccc}
\hline
\hline
Galaxy  & $B_{\mathrm{tot}}$   & $B_{\mathrm{ran}}$    & $B_{\mathrm{ord}}$   & $B_{\mathrm{tot}}$ & $B_{\mathrm{ran}}$ & $B_{\mathrm{ord}}$ \\
        & 4.85 GHz     & 4.85 GHz     & 4.85 GHz    & 8.35 GHz  & 8.35 GHz  & 8.35 GHz  \\
\hline
NGC2976 & $5.7\pm0.8$ & $ 5.5\pm0.8$ & $1.5\pm0.2$ &    $-$    &    $-$    &   $-$     \\
NGC3239 & $6.9\pm0.9$ & $ 6.8\pm0.9$ & $0.9\pm0.1$ & $7.3\pm0.9$ & $ 7.1\pm0.9$ & $<2.1$ \\
NGC4027 & $9.0\pm1.3$ & $ 8.9\pm1.2$ & $1.3\pm0.2$ & $8.1\pm1.2$ & $ 7.9\pm1.2$ & $<1.9$ \\
NGC4605 & $6.4\pm0.9$ & $ 6.2\pm0.8$ & $<1.6$ &    $-$    &    $-$    &   $-$     \\
NGC4618 & $6.0\pm0.9$ & $ 5.9\pm0.8$ & $1.3\pm0.2$ & $6.8\pm0.9$ & $ 6.6\pm0.9$ & $<1.6$ \\
NGC5204 &      $-$    &    $-$       &   $-$       & $6.3\pm0.9$ & $ 5.9\pm0.8$ & $<2.2$ \\
UGC11861& $5.4\pm0.8$ & $ 5.3\pm0.8$ & $1.2\pm0.2$ &    $-$    &    $-$    &   $-$     \\
\hline
\end{tabular}
\label{t:magnet}
\end{table*}

The radio emission of galaxies is comprised of synchrotron and thermal (free-free)
components. Having the both components separated, the synchrotron emission can be used
for estimation of strength of magnetic fields, provided energy equipartition between
magnetic fields and cosmic rays is maintained (Beck \& Krause \cite{beck05}). The best
method to estimate the thermal component consists in modelling the nonthermal spectral
index and the level of thermal fraction based on multi-frequency radio measurements.
However, we cannot apply this method for our galaxies as there are not sufficient
(literature) data concerning their radio measurements.

In our previous investigations of dwarf galaxies within the Local Group (Chy\.zy et
al. \cite{chyzy11}), we determined thermal fractions from H$\alpha$ fluxes, assuming
that dust attenuation is negligible. While justified for dwarf galaxies, this assumption
may not be valid for more massive, peculiar and Magellanic-type
objects. To mitigate the effect of dust absorption, we apply a method here of
correcting the observed H$\alpha$ fluxes for dust attenuation,
using information on the infrared (dust) emission. Such a method was proposed
by Calzetti et al. (\cite{calzetti07}) to construct the most robust galactic SFR
indicators, based on the observed H$\alpha$ and $24\,\mu$m luminosities. Having
such calibration factors of the SFR (Kennicutt \& Evans \cite{kennicutt12}, Calzetti
et al. \cite{calzetti07}) and unobscured H$\alpha$ fluxes at hand, we combined them with the
theory of thermal emission (Caplan \& Deharveng \cite{caplan86}, Niklas et al.
\cite{niklas97}). This led us to a formula for the radio thermal emission for a galaxy
at a given frequency $\nu$:
\begin{eqnarray}
 F_{\nu}({\rm{mJy}})=2.238\times 10^9  \left(\frac{F_{\mathrm{H}\alpha} + 0.020\,\nu_{\mathrm{IR}}\,F_{\mathrm{IR}}}
{\rm{erg s^{-1} cm^{-2}}}\right) \times \left(\frac{T_\mathrm{e}}{K}\right)^{0.42} \times \nonumber \\
\times \left[\ln\left({0.04995\over{\nu(\rm{GHz})}}\right)+1.5\,\ln\left(\frac{T_\mathrm{e}}{K}\right)\right],
\end{eqnarray}
\noindent where $F_{H\alpha}$ is the observed H$\alpha$ flux, $F_{\mathrm{IR}}$ is
the infrared flux at $24\,\mu$m (or $25\,\mu$m) and $T_{\mathrm{e}}$ denotes
the temperature of thermal electrons.

To calculate the radio thermal emission for our galaxies through the method described,
we used the observed H$\alpha$ fluxes from Das et al. (\cite{das12}) for NGC\,4027, James et al.
(\cite{james04}) for UGC\,11861, and Kennicutt (\cite{kennicutt08}) for the rest of galaxies.
The infrared fluxes at $24\,\mu$m were taken from SPITZER (Dale et al. \cite{dale09}) or
at $25\,\mu$m from IRAS (Sanders et al. \cite{sanders03}). In our calculations we assumed  that
$T_{\mathrm{e}}\approx 10^{4}$\,K for all the objects involved.

The obtained thermal fractions at 4.85\,GHz (see Table \ref{t:radio4}) range from 0.11
(for UGC\,11861) to 0.34 (for NGC\,3239). These values are similar to those for massive
spiral galaxies but likely lower than for dwarfs. For the dwarfs IC\,10 and NGC\,6822,
we derived thermal fractions at 4.86\,GHz of about 0.5-0.6 (Chy\.zy et al.~\cite{chyzy03}).
At 8.35\,GHz, the thermal fractions for our galaxies are larger (Table \ref{t:radio8}) and reach
0.5 for NGC\,3239.

Having estimated the thermal radio emission, we separated the synchrotron
component from the total radio fluxes at both frequencies.
The total synchrotron emission is related to the total magnetic field,
which contains two main components: the random and the ordered (uniform) one.
Information on the ordered component comes from the polarised emission detected
or from the upper limit of the polarised signal. Such fields can originate
either from compression/shearing of turbulent random fields or from
the large-scale dynamo process (Hanasz et al. \cite{hanasz09}, Gressel
et al. \cite{gressel08}).

On the basis of synchrotron component of the radio emission, we estimate the
equipartition magnetic field strengths for all our objects (Table \ref{t:magnet}).
In our calculations (cf. Beck \& Krause \cite{beck05}), we assumed a constant
ratio of protons to electrons $K=100$, a nonthermal spectral index of 0.8,
and an unprojected synchrotron disc thickness of $L=1\,$kpc. We adopted the
inclination angle from published data (Table \ref{t:sample}) to calculate the
effective synchrotron pathlength.

From our radio data at 4.85\,GHz, the strengths of the galactic total fields
averaged over galaxies are within 5.4-9.0\,$\mu$G, while those of the ordered field
are within 0.9-1.5\,$\mu$G (Table \ref{t:magnet}). The data at 8.35\,GHz lead to
a similar conclusion, as they yield magnetic field strengths similar to those for 4.85\,GHz
within the estimated uncertainties (Table \ref{t:magnet}). Hence, the values of
total and ordered field strengths are both smaller than in typical spiral galaxies, where
their values are about 10\,$\mu$G and 5\,$\mu$G, respectively (Beck \& Wielebinski \cite{beck13}).
They are both larger than in typical dwarfs of the Local Group, for which
the mean total field strength is $<4.2\,\mu$G, and the ordered field strength is $<0.9\,\mu$G
(Chy\.zy et al. \cite{chyzy11}). Therefore, by considering magnetic properties,
Magellanic-type galaxies constitute an intermediate class of objects between dwarfs
and massive spirals.

The observed magnetic fields were strongest for NGC\,4027 and NGC\,3239,
which are weakly interacting (Sect. \ref{s:radio}), and weakest
for UGC\,11861, which is an isolated galaxy. The difference (about 3\,$\mu$G)
is small, indicating that the influence of gravitational
interaction on the magnetic field strength is rather feeble for these objects. This agrees with the
results of Drzazga et al. (\cite{drzazga11}), showing that magnetic fields are only
strongly enhanced for major merger systems, especially for those undergoing the phase of
coalescing cores.

\subsection{Magnetic fields and SFRs}
\label{s:magneticSFR}

\begin{figure}[top]
\includegraphics[angle=0,width=\columnwidth]{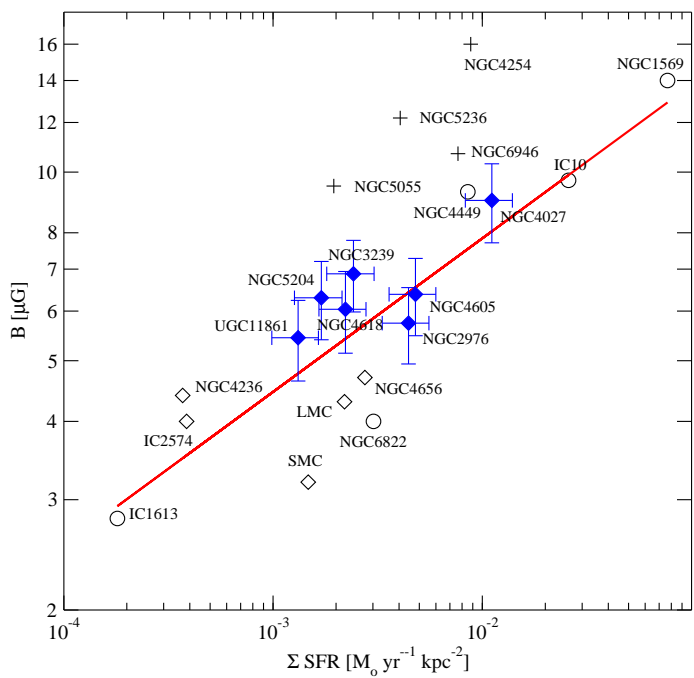}
\caption{Total field strength against surface density of the star
formation rate $\Sigma$\,SFR for the sample of galaxies of various types.
Dwarfs -- circles, Magellanic and peculiar
low-mass galaxies -- diamonds, solid diamonds -- objects presented in
this paper, and crosses -- spiral galaxies. The solid line presents the power-law
fit to 17 dwarfs and Magellanic-type objects. 
}
\label{f:b_sigmasfr}
\end{figure}

For the purpose of a more detailed and statistically valid study of magnetic properties
of low-mass objects, we enlarged our sample by including a number of cases known from
the literature. Thus, we increased the number of Magellanic-type galaxies by five: three
from our previous studies (NGC\,4236, NGC\,4656, and IC\,2574;
Chy\.zy et al. \cite{chyzy07}) and two Magellanic Clouds (LMC: Gaensler et
al.~\cite{gaensler05}; SMC: Mao et al.~\cite{mao08}). As we aim to verify the relations
found for small dwarf galaxies from our Local Group (Chy\.zy et al. \cite{chyzy11}),
we include those LG dwarfs for which radio emission had been detected
(IC\,1613, NGC\,6822 and IC\,10) in the analysis.  We add two starburst galaxies with
well-defined radio properties: NGC\,1569 (Kepley et al.~\cite{kepley10}) and
NGC\,4449 (Chy\.zy et al. \cite{chyzy00}).  In effect, the enlarged sample includes
ten Magellanic-type galaxies, two peculiar galaxies and five dwarf irregulars, thus 17 objects in total.
For a rough comparison with spiral galaxies we also analyse four massive spirals
($M_{\mathrm{HI}} > 4\times 10^9\,\rm{M}_{\sun}$): NGC\,5055, NGC\,5236, NGC\,6946
from the compilation of Basu \& Roy (\cite{basu13}) and NGC\,4254 from our previous work with Chy\.zy
et al. (\cite{chyzy07}). These spirals represent the Hubble morphological types c and d.

We constructed a diagram showing the relation of total magnetic field strengths $B$
and (surface) densities of the SFR ($\Sigma$\,SFR) for all galaxies (Fig. \ref{f:b_sigmasfr}).
The SFRs for seven galaxies from our main sample were estimated from the SFR
indicator of Calzetti et al. (\cite{calzetti07}), as mentioned in the previous Section. For
other galaxies from the extended sample, the SFRs were taken from the papers mentioned above.
The global SFRs were then divided by the observed optical areas of galaxies
(Table \ref{t:sample}) and rescaled to give galaxy-mean $\Sigma$\,SFRs in
$\rm{M}_{\sun}\, \rm{yr}^{-1}\, \rm{kpc}^{-2}$.
The observed Magellanic-type galaxies have total magnetic field strengths and $\Sigma$\,SFR
between the radio-weakest dwarf galaxy IC\,1613 and the starburst dwarfs NGC\,1569 and IC\,10
(Fig. \ref{f:b_sigmasfr}). Magnetic fields in spirals are as strong as in starburst dwarfs and
relatively stronger than in Magellanic-type galaxies.
For the extended sample of 17 objects, the strength of the total field $B$
is strongly related with $\Sigma$\,SFR, manifesting a power-law relation with an
exponent $a=0.25\pm0.02$. This relation is statistically highly significant with
a correlation coefficient $r =0.87$, and the significance level (p-value of null
hyphothesis) $P=5\times10^{-6}$. It confirms $B-\Sigma$\,SFR relation found
for a small sample of Local Group dwarfs with the power-law exponent
of $a=0.30\pm0.04$ (Chy\.zy et al. \cite{chyzy11}). In the present  analysis
the statistical sample is more than twice as large than the previous one and
not only supports but also even strengthens the results obtained from the Local Group dwarfs.

\begin{figure}[top]
\includegraphics[angle=0,width=\columnwidth]{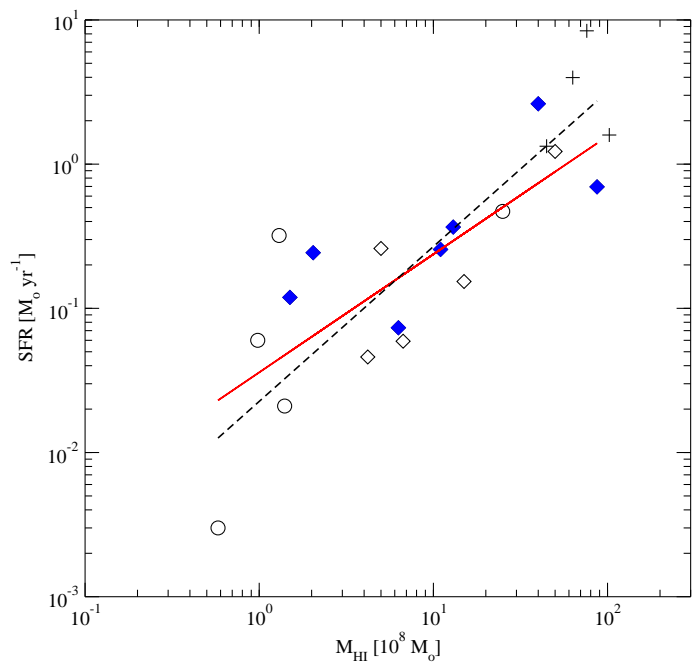}
\caption{Correlation of global star formation rate with \ion{H}{I} mass for dwarf
galaxies (circles), Magellanic-type and peculiar galaxies (diamonds) and normal spirals (crosses).
Solid diamonds mark objects presented in this paper. Solid and dashed lines demonstrate the
classical Y-X regression fit and the bisector fit to 17 Magellanic-type and dwarf galaxies.
}
\label{f:m_sfr}
\end{figure}

A similar phenomenon of magnetic field production increasing with the star-forming activity
in galaxies was also found locally within individual objects -- NGC\,4254 with the
exponent $a=0.18\pm0.01$ (Chy\.zy et al. \cite{chyzy08}) and NGC\,6946 with $a=0.14\pm0.01$
(Tabatabaei et al. \cite{tabatabaei13}). The differences in these slopes could have arisen
from different contents of the ordered field and/or galaxy environment (Tabatabaei et al.
\cite{tabatabaei13}, Krause \cite{krause09}).

We note a substantial scatter of points of about 0.5 dex to be seen in the $B-\Sigma$\,SFR
diagram, which could obscure such a relation if the range of values of $\Sigma$\,SFR was
smaller. In the present study, the pattern is distinct only when almost three
orders of magnitude in $\Sigma$\,SFR (Fig. \ref{f:b_sigmasfr}) are considered.
A significant spread in the global properties of galaxies is therefore necessary to
disclose $B-\Sigma$\,SFR relation.

According to our analysis, we also note that  {\em global} SFR and the galactic \ion{H}{i}
mass ($M_{\mathrm{HI}}$) are correlated ($r=0.74$, $P=10^{-4}$). The estimated power-law slope
for this relation $a=0.82 \pm 0.20$ from the classical regression fit (Fig. \ref{f:m_sfr})
and $a=1.07 \pm 0.16$ from the bisector fit.  The \ion{H}{i} mass is strongly correlated
with the total galactic mass ($M_{\mathrm{tot}}$, see Sect. \ref{s:sample}), $r=0.92$, for 
our sample. Hence, a similar relation holds between the SFR and the total mass with 
$a=0.80\pm 0.26$. A similar dependency with a slope of $\approx 1.0$  for dense molecular 
gas of spiral galaxies was found by Gao \& Salomon (\cite{gao04}). Slightly steeper
($a=1.5$) relations were estimated for the \ion{H}{i} mas of galaxies of various
morphological types in the Local Volume (Karachentsev \& Kaisina \cite{karachentsev13})
and for dwarf galaxies in the Local Group ($a=1.4$, Chy\.zy et al. \cite{chyzy11}).
There are some observational and theoretical hints, in which  linearly large low surface
brightness galaxies can show lower SFR than the high surface brightness galaxies for the
same reservoir of neutral hydrogen (Boissier et al. \cite{boissier08}). Consequently,
differences in the star formation efficiency could cause differences in SFR-mass relations
among different samples and also enlarge the spread in the observed  $B-\Sigma$\,SFR relation
in our sample.

\section{Discussion}
\label{s:discussion}

\subsection{Radio--FIR correlation}

\begin{figure*}
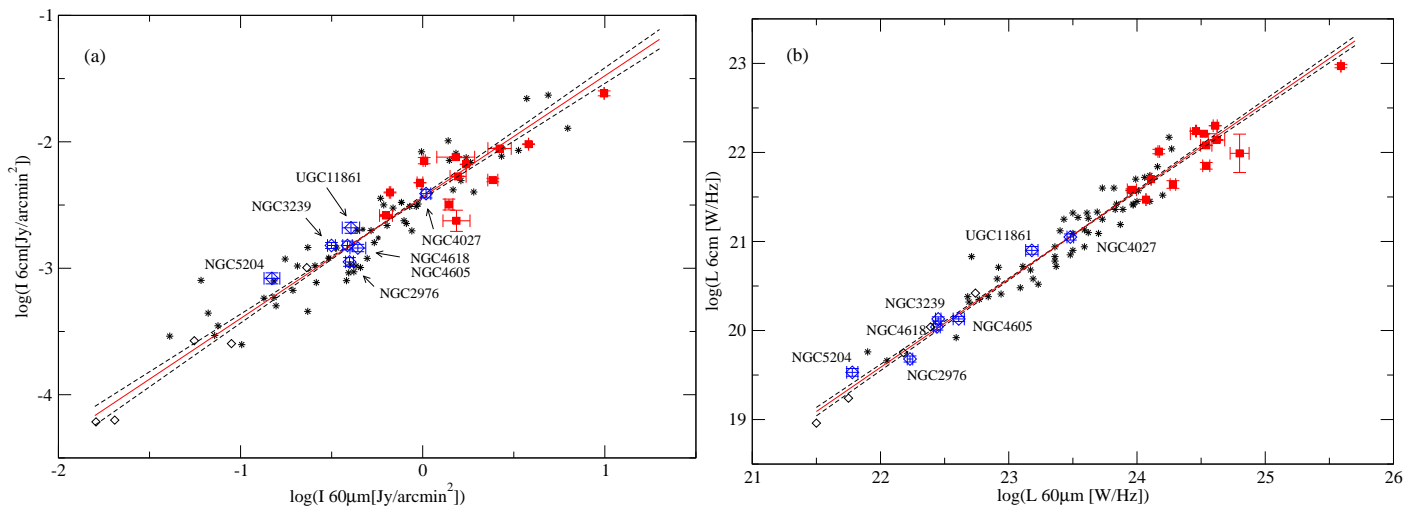

\includegraphics[angle=0,width=9.09cm,clip]{radiofir.eps}
\includegraphics[angle=0,width=9.2cm,clip]{radiofirpower.eps}
\caption{Radio--FIR correlation diagram for our Magellanic-type and
peculiar galaxies (diamonds), as well as interacting objects (squares) and bright spiral
galaxies (asterisks). The surface brightness (a) and luminosity (b) at 4.85\,GHz (6\,cm) 
and at 60\,$\mu$m  are used, respectively. The solid line is a bisector fit to all galaxies,
while the dashed lines represent simple X vs. Y and Y vs. X regressions. Error
bars are marked just for a few objects for example's sake.
}
\label{f:radio_fir}
\end{figure*}

\begin{table*}[t]
\caption{Power-law fits to radio -- infrared relations based on 
surface brightness and luminosity of 79 galaxies of various types.}

\centering
\begin{tabular}{lcccc}
\hline
\hline
Fit   & \multicolumn{3}{c}{Slopes}  & Correlation \\
      & radio-FIR & FIR-radio & radio-FIR  \\
      & Y-X       & X-Y       & bisector   \\
\hline
a) Surface brightness & $0.92\pm0.04$ & $ 1.01\pm0.04$ & $0.96\pm0.03$ & $0.95$  \\
b) Luminosity         & $0.97\pm0.02$ & $ 1.02\pm0.02$ & $0.99\pm0.02$ & $0.98$  \\
\hline
\end{tabular}
\label{t:radiofir}
\end{table*}

The observed connection of the magnetic field strength and the star-forming activity
in galaxies can be investigated through a relation of radio and far-infrared (FIR)
emission. The rates of local production of magnetic fields and CRs are tightly connected
to supernova explosions of massive stars. The regions responsible for vivid star formation
produce UV-photons that, in turn, cause dust heating and re-emission in the infrared band.
To investigate this relation for our low-mass galaxies, we constructed a mean
surface brightness radio--FIR diagram  using the total 6\,cm and 60\,$\mu$m emission in the radio and
infrared bands, respectively (Fig. \ref{f:radio_fir}a). We also included 
bright spirals from Gioia at al. (\cite{gioia82}) in the diagram and interacting objects from Drzazga et al.
(\cite{drzazga11}) by building up a sample of 79 objects in total. As surface brightness does
not depend on galaxy distance, it is an alternative way to present the radio--FIR
correlation instead of luminosity. For comparison's sake, we also present {\em classical}
radio--FIR relation for luminosities (Fig. \ref{f:radio_fir}b). In this case, statistical 
dependency might partly arise from an obvious relation that larger galaxies are more luminous in both  radio
and infrared bands. The obtained slopes in the power-law fits and correlation coefficients 
are shown in Table \ref{t:radiofir} for both kinds of relations.

Our investigation shows that Magellanic-type and peculiar galaxies do not deviate in any
systematic way from the power-law fit constructed for all galaxies, including interacting 
objects (Drzazga et al. \cite{drzazga11}). This statement is valid for the relation 
constructed for surface brightness (which shows a slope of 
$0.96\pm0.03$ from the bisector fit) and for luminosity (with a slope of $0.99\pm0.02$). 
The radio-FIR relation for luminosity is tighter: regression lines are closer to each other and 
the correlation coefficient is slightly larger ($r=0.98$) than for the surface brightness relation 
($r=0.95$), as expected (Table \ref{t:radiofir}).

According to the above argumentation, the radio--FIR correlation is thought
to be based on the connection between total (mostly turbulent) magnetic fields and
star-formation activity, as shown in Fig.~\ref{f:b_sigmasfr}. Therefore, a small-scale field
amplification producing random magnetic fields must work in the Magellanic-type galaxies, 
likewise, as in more massive and interacting objects.

We note that the slopes of the constructed relations were obtained for the total radio emission, which
is a  sum of thermal and nonthermal ones. Careful studies of local radio--FIR relation
in individual galaxies revealed slightly steeper slopes for thermal radio emission
and flatter slopes for synchrotron emission (e.g. Hoernes \& Berkhuijsen \cite{hoernes98}, Hughes et al.
\cite{hughes06}, Tabatabaei et al. \cite{tabatabaei13}).
The differences in slopes are probably caused by CRs propagation effects,
properties of the magnetic field, and the existence of a thick disc/halo in galaxies (Berkhuijsen et al.
\cite{berkhuijsen13}).

 The low-emission objects are of particular interest in the constructed radio--FIR relations.
For a large sample of galaxies Yun et al.~(\cite{yun01}) noticed a systematically diminishing
radio emission (at 1.4\,GHz) for objects of the lowest infrared luminosity.
These galaxies
showed a systematic tendency to lie below the best-fit line of the radio--FIR relation.
Although our sample contains low-mass objects manifesting weak magnetic field strength,
no distinct radio deficiency at 4.85\,GHz can be seen in the radio--FIR relations 
(Fig. \ref{f:radio_fir}). We therefore estimated radio luminosities at 1.4\,GHz for our 
galaxies based on the NVSS survey to check their relation to Yun et al.'s outliers. It turned 
out that our galaxies have luminosities starting from $5\times 10^{19}$\,W\,Hz$^{-1}$ 
 at 1.4\,GHz and hence are located just above the break in the radio--FIR relation provided by Yun at al. We can still conjecture that deviations from the general radio--FIR relation are to be expected
for our low-mass objects at much lower radio frequencies due to various
processes shaping low-energy CRs spectra. A statistical study of such processes will
soon be possible with all-sky LOFAR surveys (R{\"o}ttgering et al. \cite{rottgering11}).

We can thus conclude that the low-mass stellar systems of Magellanic or peculiar morphology
in our local neighbourhood feature similar physical conditions as star formation,
magnetic field, and cosmic-ray generation processes as those observed for massive or
interacting galaxies.

\subsection{Production of magnetic energy}

\begin{figure}
\includegraphics[angle=0,width=8.5cm]{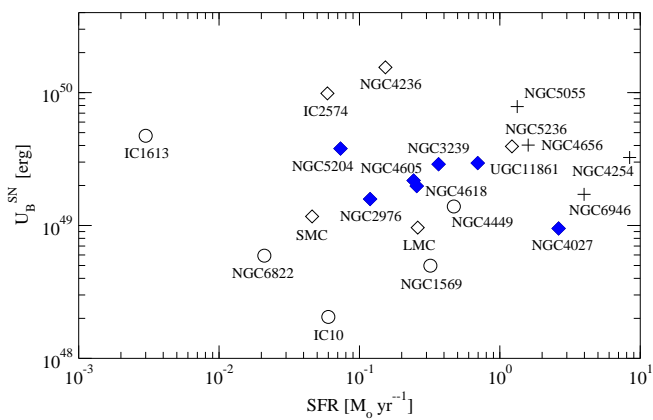}
\caption{Production of magnetic energy per individual supernova event $U^{\mathrm{SN}}_{\mathrm{B}}$ for
dwarf (circles), Magellanic-type and peculiar galaxies (diamonds) and normal spirals
(crosses) of different SFR.
The objects presented in this paper are marked with solid diamonds.
}
\label{f:sfr_eb}
\end{figure}

The radio--FIR relation reveals some universal physical processes at work in the galactic
ISM on small and large scales. Although this relation is not fully accounted for
(e.g. Roychowdhury \& Chengalur \cite{roychowdhury12}, Schleicher \& Beck \cite{schleicher13}),
one can expect that the observed (correlated) different levels of galactic radio and
infrared emission result from different rates of supernova explosions.
However, it is not well known yet if the production of magnetic field energy {\em per supernova
event} is independent from various properties of galaxies, such as the global SFR,
galactic mass, or morphology. The supernova energy release is an internal process and should
not depend on the environment. However, the production of magnetic fields is the result of
a chain of physical processes whose efficiencies may depend on
local conditions in the ISM (see e.g. Gent et al. \cite{gent13}).
For example, the turbulence driven by supernova, which by the fluctuation dynamo
produces magnetic fields, is the process intrinsically related to the ISM.
The magnetic energy production is also related to the production of CRs (cf. equipartition
assumption) which results from local acceleration processes in the diffusive shock of the
supernova remnant. The MHD simulations also suggest that fluctuation of the SFR may influence the dynamo
action and magnetic fields (Hanasz et al. \cite{hanasz04}). The enhancement of
magnetic fields on large scales also depends on galactic rotation and galactic mass.
Hence, the problem of the net production of magnetic field per supernova explosion is
an important issue for our understanding of generation mechanisms of magnetic fields in
various types of galaxies and for building up reliable MHD models of galactic evolution.

We investigate this issue for all objects in our extended sample of dwarfs, Magellanic,
peculiar, and spiral galaxies.
We calculated a mean value of the energy (volume) density of magnetic fields $u_{\mathrm{B}}$
for each studied galaxy from the estimated total magnetic field strength: $u_{\mathrm{B}}=B^2/(8\pi)$.
Contrary to the synchrotron emission and the surface
density of the SFR, it is worth to notice here that, the magnetic field strength is a purely local value, which is not integrated
over the galactic pathlength (cf. Sect. \ref{s:magneticSFR}).

The galactic supernova rate $\nu_{\mathrm{SN}}$ depends on the stellar initial mass
function (IMF). In the case of the Salpeter IMF, it takes a form of a single power-law with
an index $\gamma=2.35$. Following Kennicutt (\cite{kennicutt98}) for the low- and
high-mass limits of the IMF, we apply 0.01\,M$_{\sun}$ and 100\,M$_{\sun}$,
respectively. Only stars more massive than 8\,M$_{\sun}$ finish their evolution as supernovae,
depositing large amounts of energy in the ISM, which produces CRs in their shocks and enhancing
magnetic fields. Under these conditions, the global rate of the supernova explosions (per year)
scales for a given galaxy with the star formation rate as $\nu_{\mathrm{SN}}=0.02\,$SFR.

\begin{figure}[t]
\includegraphics[angle=0,width=8.5cm]{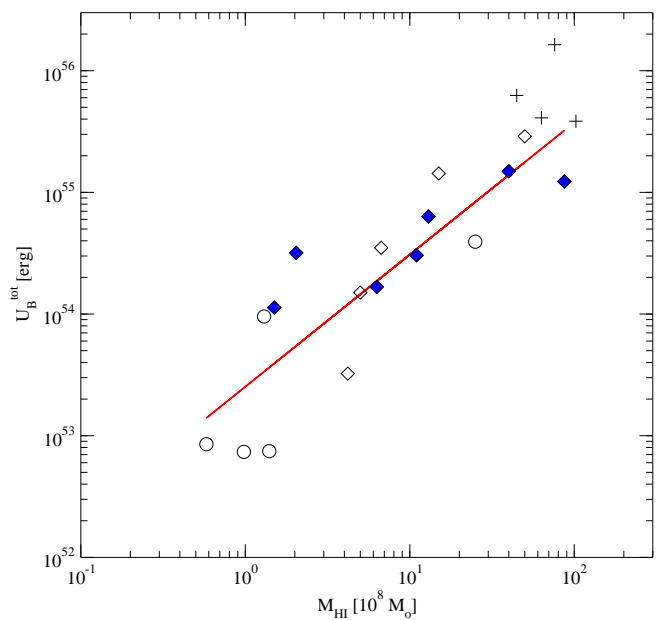}
\caption{Total magnetic energy $U^{\mathrm{tot}}_{\mathrm{B}}$ versus
galactic \ion{H}{i} mass. Circles -- dwarf galaxies, diamonds -- Magellanic-type and peculiar galaxies,
crosses -- normal spirals, and solid diamonds -- the objects are presented in this paper.
The solid line presents the power-law fit to the extended sample of 17 dwarfs and
Magellanic-type objects.
}
\label{f:m_ebtot}
\end{figure}

The observed synchrotron emission of galaxies results not only from the very recent supernova
explosions described by the current SFR but also from the earlier ones,
whose magnetic fields have not dissipated yet by turbulent diffusion. A crude estimation of this time
can be made from a characteristic diffusion time scale $\tau=L^2/\eta$, where $\eta=1/3\,{\mathrm v}L$
is the turbulent magnetic diffusivity, $\mathrm{v}$ is the characteristic turbulent velocity, and
$L$ is the magnetic field eddy scale. Hence, the magnetic energy produced per supernova event is
\begin{equation}
 U^{\rm{SN}}_{\mathrm{B}}=\frac{u_\mathrm{B}\,V}{\nu_{\rm{SN}}\,\tau}=\frac{B^2\,a^2 h}{8\,\nu_{\rm{SN}}\,\tau},
\end{equation}
where $V=\pi a^2 h$ is the approximate galaxy disc volume with a radius $a$ and
a scale height $h$. In our approximations for the galactic radius, we apply the linear
size of the major axis (Table \ref{t:sample}) and a value of 1\,kpc or 0.5\,kpc for the height $h$,
depending on whether the observed galactic minor axis is larger or smaller than 1\,kpc. We assume
the turbulent velocity of the ISM $\mathrm{v}\approx 10$\,km\,s$^{-1}$ and approximate the magnetic
field scale in the ISM as $L\approx 100$\,pc, which gives $\eta \approx 10^{26}\rm{m}^2\,\rm{s}^{-1}$.

The obtained diagram, which shows magnetic energy per supernova event $U^{\mathrm{SN}}_{\mathrm{B}}$
across different levels of global SFR, is presented in Fig.~\ref{f:sfr_eb}. There is no
specific relation to be seen (the estimated slope is $-0.04\pm 0.13$), and the correlation
coefficient is not significant either ($r=-0.07$, $P=0.77$). The roughly constant
magnetic field production per supernova event for a variety of objects
from dwarfs and Magellanic-type objects to massive galaxies, actually suggests that
the same processes are at work in the ISM in all these objects. In this scenario,
starburst dwarf galaxies and large, massive galaxies can produce stronger magnetic
fields just because supernova explosions are denser in these objects in time or in space.
For example, the slowest forming-stars object in the analysed sample, IC\,1613, shows
the same production of magnetic energy ($4\times10^{40}$\,erg) as NGC\,4656, which has
two and half orders of magnitude larger star formation rate.

 Close inspection of the relation also gives some hints that among galaxies of the same SFR, like
  IC\,2574 and IC\,10, those which are larger in size (like IC\,2574) produce larger magnetic energy density per
single supernova event. The same is observed for several other galaxies. If it
is not caused by a selection effect, this would suggest that the most effective production of magnetic fields energy density occurs 
in a less compact star-forming regions and in larger galaxies of low-surface brightness.
Indeed, the power-law fit shows weak dependence $U^{\mathrm{SN}}_{\mathrm{B}}\propto (\mathrm{SFR}/V)^{-0.58\pm0.06}$
with high negative correlation ($r=-0.9$). This effect could be caused by a stronger
diffusion of CR electrons or/and stellar winds from more compact star formation regions
with clustered supernovae. This effect also resembles the results of MHD modelling that
too strong star-forming activity in galaxies can quench magnetic
fields production (Hanasz et al. \cite{hanasz04}).

Our calculations also provide an independent estimation of the value of average magnetic
energy release per supernova event: $E^{\rm{SN}}_{\mathrm{B}} \approx 3\times10^{49}$\,erg
(Fig. \ref{f:sfr_eb}). If the supernova explosion exhausts the canonical total kinetic energy of
about $10^{51}$\,erg, then the production of magnetic energy is at the level of about
3\% of this total energy. This is an independent estimation of the parameter that needs
to be assumed in MHD simulations of magnetic field evolution in galaxies
(e.g. Hanasz et al. \cite{hanasz09}, Gressel et al. \cite{gressel08}).

Furthermore, we compared total magnetic field energy in galaxies using the equation
$U_{\mathrm{B}}^{\mathrm{tot}}=u_{\mathrm{B}}\, V$ and galactic
\ion{H}{i} mass (Fig. \ref{f:m_ebtot}). In this case,
a strong power-law relation (with an index of $1.09\pm0.16$) is clearly discernible and
quantitatively confirmed  by the large correlation coefficient $r=0.86$ ($P=9\times10^{-6}$).
This relationship can be divided by the galactic volume, which yields
an approximate dependence of magnetic field strength on the local \ion{H}{i}
gas density: $B\propto \rho_{\mathrm{HI}}^{0.54}$.
We checked this relationship for the total gas masses of galaxies $M_{\rm gas}$, which
includes both the atomic and molecular gas phases, as determined by the simple relation
$M_{gas}=1.3\times(M_{\mathrm{HI}}+M_{\mathrm H_2})$ (Boselli \cite{boselli12}).
For 10 out of 17 galaxies, we found H$_2$ masses in the literature. For the rest
(NGC\,3239, NGC\,4027, NGC\,4605, NGC\,4618, NGC\,5204, UGC\,11861, and NGC\,4656),
we estimated the $M_{gas}$ from the gas-to-dust ratio and the dust mass from
$100\,\mu$m IRAS data, according to the method outlined by Boselli (\cite{boselli12}).
For this, case the power-law fit to the $U_{\mathrm{B}}^{\mathrm{tot}}-M_{gas}$ relation gives an exponent of
$1.14\pm 0.12$ ($r=0.88$, $P=2\times10^{-7}$), which results in a $B\propto \rho_{\mathrm{gas}}^{0.57}$
relationship. The values of the exponent in the $B$--gas density relations
for both the \ion{H}{I} and the total gas agree within the statistical uncertainties
with the theoretical value of 0.5 for equipartition of the magnetic
field energy and the turbulent energy of the ISM. Such an equipartition was also
found in MHD simulations of  Cho \& Vishniac (\cite{cho00}). A slightly smaller
slope of $0.48\pm0.05$ for the total cool gas (\ion{H}{i} plus H$_2$) and the full
range of galactic morphology, including high surface-brightness and interacting
galaxies, was reported by Niklas \& Beck (\cite{niklas97}).

The $B-$gas density relation represents a portion of processes leading to the
observed radio--FIR correlation. In addition, the correlations involve other
processes, like coupling between magnetic fields and CRs, the production of CRs
(cf. equipartition assumption), the Schmidt-Kennicutt law, and dust re-radiation
of UV photons, giving far-infrared indicators of the SFR (Basu et al.
\cite{basu12}). All the presented results prove that the low-mass galaxies from
the analysed sample show several common patterns for whole galaxies. To better
understand these processes, a future analysis of more massive galaxies is desirable
to cover a wider span of analysed parameters and to allow multi-dimensional
statistical analysis.

\section{Conclusions}
\label{s:conclusions}

We performed radio polarimetric observations of seven low-mass galaxies with the Effelsberg 100-m
telescope at 4.85\,GHz and/or 8.35\,GHz. Five galaxies are classified as Magellanic-type and
two (NGC\,2976 and NGC\,4605) represent a class of peculiar ``pure disc'' objects with low mass.

We found the following:

\begin{itemize}

\item
The observations of all seven galaxies indicate that their radio emission closely follow optical properties
of the discs, regardless of whether galaxies are isolated or show signs of tidal interactions.

\item {For the observed galaxies, radio thermal fractions range from 0.11 to 0.34}
at 4.85\,GHz and correspond more to large spirals than to
dwarf galaxies. Similarly, radio spectral indices with a mean of about 0.67 are
closer to spiral galaxies ($\approx 0.74$) than to dwarf objects ($\approx 0.38$).

\item The estimated strengths of the total magnetic field is in the range of
$5-9\,\mu$G and those of the ordered fields $1-2\,\mu$G. They are significantly
weaker than for typical spirals. At 4.85\,GHz, the polarised emission was detected for
five out of six galaxies, showing a small polarisation degree of only 1-4\%.

\item  Our analysis of the extended sample of 17 galaxies shows that the Magellanic-type
galaxies fit well into the pattern obtained for dwarf galaxies in the Local Group
(Chy\.zy et al. \cite{chyzy11}), indicating that the production of
turbulent magnetic field is related to the surface density of the SFR with
a power-law index of $0.25\pm 0.02$. Similar patterns of magnetic field production
were also found to work locally within single galaxies (Tabatabaei et al. \cite{tabatabaei13},
Chy\.zy et al. \cite{chyzy08}).

\item The Magellanic-type galaxies from our extended sample hold the general radio (4.85\,GHz)
and far-infrared (60\,$\mu$m) power-law relation, which is determined for surface brightness of galaxies of 
various types with a slope of $0.96\pm0.03$ and correlation coefficient $r=0.95$. 
The ascertained far-infrared relation based on luminosity of galaxies is tighter and steeper (with a slope of 
$0.99\pm0.02$ and $r=0.98$) but likely includes a partial correlation from a simple tendency of larger 
objects, which is also more luminous in both radio and infrared bands.

\item The estimated production of magnetic field energy per supernova event in
low-mass galaxies does not depend on either  the galactic global SFR ($r=-0.07$, $P=0.77$),
 on morphological type, or on galactic mass and is roughly constant for all the objects.
We estimated magnetic energy release as about 3\% ($3\times 10^{49}$\,erg)
of the total available kinetic energy in an individual supernova explosion (about
$10^{51}$\,erg).

\item The total magnetic field energy of our galaxies scales almost linearly with the galactic
\ion{H}{I} mass (with the power-law index of $1.09\pm 0.16$) and the total atomic and molecular gas
mass (with the exponent of $1.14\pm 0.12$). This results in scaling of the total magnetic field strength with
the local density of \ion{H}{I} ($B_{\mathrm{tot}}\propto \rho_{\mathrm{HI}}^{0.54}$) and total \ion{H}{I} + H$_2$
gas ($B\propto \rho_{\mathrm{gas}}^{0.57}$). Such relations indicate
equipartition of the magnetic energy and the turbulent kinetic energy of the gas.
This supports the conclusion that the total magnetic field is dominated by the
turbulent component, which results from the small-scale dynamo process rather than from
the large-scale one.
\end{itemize}

\begin{acknowledgements}
We thank an anonymous referee for helpful comments and suggestions.
This research has been supported by the scientific grant from the Polish National
Science Centre (NCN), decision no. DEC-2011/03/B/ST9/01859. RB and UK acknowledge support
from DFG Research Unit FOR1254.
We acknowledge the use of the HyperLeda (http://leda.univ-lyon1.fr) and NED
(http://nedwww.ipac.caltech.edu) databases.
\end{acknowledgements}

\end{document}